\nopagebreak  
\RequirePackage{latexrelease}
\documentclass[aps,reprint,10pt,showkeys,showpacs,superscriptaddress,showpacs]{revtex4-2}
\usepackage{multirow}
\usepackage{enumitem}
\usepackage{amsfonts} 
\usepackage{amsmath}
\usepackage{amssymb}
\usepackage{graphicx}
\usepackage{subfigure}
\usepackage{color}
\usepackage{enumerate}
\usepackage[]{natbib}
\usepackage{sidecap}
\usepackage{soul}
\usepackage{cancel}
\usepackage{ulem}
\usepackage{dcolumn}%
\usepackage{bm}%

\newcommand*\xbar[1]{%
  \hbox{%
    \vbox{%
      \hrule height 0.5pt 
      \kern0.5ex
      \hbox{%
        \kern-0.1em
        \ensuremath{#1}%
        \kern-0.1em
      }%
    }%
  }%
} 

\begin{document}

\title[]{Rayleigh-Taylor instability in inhomogeneous relativistic classical and degenerate electron-ion magnetoplasmas}  
\author{Rupak Dey}
\email{rupakdey456@gmail.com }
\affiliation{Department of Mathematics, Siksha Bhavana, Visva-Bharati University, Santiniketan-731 235,  India
} 
 \author{A. P. Misra}%
 \homepage{Author to whom any correspondence should be addressed}
 \email{apmisra@visva-bharati.ac.in}
 \affiliation{Department of Mathematics, Siksha Bhavana, Visva-Bharati University, Santiniketan-731 235,  India
} 

\date{\today}

\begin{abstract}
We study the Rayleigh-Taylor instability (RTI) of electrostatic plane wave perturbations in  compressible relativistic  magnetoplasma fluids with thermal ions under gravity in three different cases of when (i) electrons are in isothermal equilibrium, i.e., classical or nondegenerate, (ii) electrons are fully degenerate (with $T_e=0$), and (iii) electrons are partially degenerate or have finite temperature degeneracy (with $T_e\neq0$). While in the cases of (i) and (iii), we focus on the regimes where the particle's thermal energy is more or less than the rest mass energy, i.e., $\beta_e \equiv k_{\mathrm B}T_e/{m_ec^2}<1~\rm{or}~>1$, the case (ii) considers from weakly to ultra-relativistic degenerate regimes. A general expression of the growth rate of instability is obtained and analyzed in the three different cases relevant to laboratory and astrophysical plasmas, which generalize and advance the previous theory on RTI.
\end{abstract}
\maketitle

\section{Introduction} \label{sec-intro}
The Rayleigh-Taylor instability (RTI) can occur at the interface between two fluids of different densities or fluids with sharp density gradients when a heavier fluid is accelerated by a lighter one under the influence of the gravitational force and the magnetic field \cite{maryam2021rayleigh,bhambhu2024radiation, garai2020rayleigh,chandrasekhar1961hydrodynamic, onishchenko2011magnetic}. In classical plasmas, the RTI growth rate gets enhanced due to the density difference of fluids by $\tilde{\gamma}=\sqrt{A_Tkg}$, where $A_T,~ k$, and $g$ are, respectively,  the Atwood number, perturbed wavenumber, and gravitational force. The  RTI can also occur in inertial confinement fusion (ICF) \cite{betti2016inertial} and several astrophysical environments, such as crab nebula \cite{tanaka2018confinement}, interacting supernova \cite{duffell2017rayleigh}, supernova remnants \cite{rigon2019rayleigh}, white dwarfs  \cite{peterson2021effects, saumon20221current},  neutron star \cite{bejger2004surface,lasky2015gravitational}, and disk magnetosphere interface \cite{zhu2023global} where the rare magnetospheric plasma provides a support to highly dense matter against the gravitational force.  Chandrasekhar \cite{chandrasekhar1961hydrodynamic} first investigated the linear dispersive properties of waves propagating at the interface between two semi-infinite fluids in a Tokamak plasma. Later, it was advanced to study the instability analysis of an electrolyte solution in the presence of a strong magnetic field \cite{yang2019autophoresis}. The theoretical developments on RTI are also noted in other plasma environments, e.g., dusty plasmas \cite{avinash2015rayleigh, dolai2016effect}, ionospheric region \cite{sekar1999effects}, dayside magnetopause \cite{yan2023rayleigh, archer2019direct}, etc. Here, the fluid viscosity and shear flow effects significantly affect the instabilities in a uniform magnetized plasma    \cite{adams2015observation,mikhailenko2002rayleigh}.    
\par
 Typically, astrophysical plasmas are highly dense and degenerate gas, and also they can be strongly magnetized.  In white dwarfs, the particle number density can vary in the range $n_i \simeq 10^{28}-10^{30}~\rm{ cm}^{-3}$, the temperature varies from $10^6$ K to $10^8$ K, and the magnetic field $\sim10^4-10^8$ G. So, the physical parameters, including the particle density, magnetic field, and the particle's temperature can play a significant role in the linear and nonlinear propagation of waves and instabilities in such plasma environments.  The degenerate electrons may be non-relativistic, relativistic, or ultra-relativistic according to when the Fermi energy is much larger than, close to, or much smaller than the electron rest-mass energy. Thus, it is pertinent to define a pressure law and the particle distribution that can efficiently describe the relevant physics of relativistic degenerate electron-ion (e-i) magnetoplasmas. In this context, several authors have studied linear and nonlinear properties of waves and instabilities in relativistic and non-relativistic plasmas \cite{bhambhu2023rayleigh, misra2018stimulated, maryam2021rayleigh, turi2022magnetohydrodynamic,chakrabarti1996velocity, chakrabarti1998rayleigh,shahmansouri2016,shahmansouri2017}.   
\par
In recent work, Maryam \textit{et al.} \cite{maryam2021rayleigh} investigated the RTI in radiative dense electron-ion (e-i) plasmas, and they showed the wave phase velocity and the instability growth rate are influenced by the radiative pressure of electrons.  Starting from a quantum hydrodynamic model, Bychkov \textit{et al.} \cite{bychkov2008rayleigh} investigated the quantum effects on the internal waves and RTI in quantum plasmas. They showed that quantum pressure always stabilizes the RTI, and a specific form of pure quantum internal wave exists in the transverse direction. This type of internal wave was absent in classical plasmas and not influenced by the gravity force. The presence of an external magnetic field beside the quantum effect stabilized the RTI more, as studied by Hoshoudy \cite{hoshoudy2009quantum}. Ali \textit{et al.} \cite{ali2009rayleigh} studied the RTI in compressible quantum e-i magnetoplasmas with cold classical ions and inertialess electrons, and they showed that the quantum speed and density gradient scale length enhanced the growth rate of instability due to quantum correction associated with the Fermi pressure law and quantum Bohm potential force. In a similar study, Adak \textit{et al. }  \cite{adak2014rayleigh} reported the RTI in a pair-ion (positive and negative ions with equal masses but different temperatures)
inhomogeneous plasma under the influences of the gravity force and the density gradient. They showed that the formation of RTI neither depends on the ion masses nor the growth rate of instability. A development in RTI has also been noted in a strongly coupled e-i quantum plasma under the influence of the shear viscosity, and the growth rate destabilizes irrespective of the direction of the gradient of shear viscosity \cite{garai2020rayleigh}. Later, Bhambhu and Prajapati \cite{bhambhu2023rayleigh} studied the effects of compressibility,  ultra-relativistic degenerate electrons, and strong coupling effect of classical ions on the density gradient-driven RTI  opposite to the direction of gravity in unmagnetized plasmas. They showed that the Coulomb coupling parameter and isothermal compressibility of ion fluid significantly modify the growth rate of RTI in plasmas relevant to white dwarfs. In contrast, the ultrarelativistic degenerate pressure of electrons became insignificant in growth rate but significantly modified the existing regime of RTI; however, they did not consider the relativistic monition of electrons and ions and the finite temperature degeneracy of electrons. After that, authors studied the dispersion properties of RTI analytically in kinetic and hydrodynamic regimes with the effects of radiation pressure of ultra-relativistic degenerate electrons and strong magnetic field \cite{bhambhu2024radiation}. The viscosity and resistivity effects in low magnetized and high energy density e-i solar plasmas have been investigated numerically for high Atwood numbers on the formation of RTI \cite{Bera2022the}. The growth rate drastically changes due to adding the viscosity and resistivity effects as well as its internal structure to a small-scale limit. The magnetic field is explicitly independent of viscosity, and enstrophy and kinetic energy are independent of resistivity. More advancements in RTI and internal waves in strongly coupled in-compressible quantum rotating plasmas can also be found \cite{khan2023investigation}. Thus, in the current situation, RTI is an exciting problem for the researcher due to its several applications in laboratory plasmas and astrophysical environments.
\par 
 From the above investigations,  none of the authors have considered the effect of relativistic plasma fluid motion, fully degenerate, and partially degenerate pressures of electrons for advancing the existing theory of RTI in strongly magnetized e-i plasmas. Such advancement of RTI with analytical and numerical analyses will help better understand the observed phenomena in many degenerate astrophysical objects and laboratory plasmas. In this work, we study the possibility of the onset of RTI in a compressible inhomogeneous relativistic electron-ion magnetoplasma fluids under the influences of a strong uniform magnetic field, density-gradient, and gravity force in three different cases when (i) electrons are classical or nondegenerate, (ii) electrons are fully degenerate, i.e., form a zero-temperature Fermi gas, and (iii) electrons are partially degenerate with Fermi energy higher than the electron thermal energy. Considering a two-fluid relativistic model for electrons and ions, we derive a general dispersion relation and analyze it for the RTI in the above three cases. The results show that the instability growth rate significantly differs as one switches from classical nondegenerate to degenerate regimes. We organize the manuscript in the following fashion, where we describe the physical model and the basic equations for the electron and ion fluids in Sec. \ref{sec-model}. Section \ref{sec-linear} demonstrates the linear theory of RTI and analyses the instability growth rates in three different regimes of classical, fully degenerate, and partially degenerate electrons. Finally, Sec. \ref{sec-conclusion} is left to summarize and conclude the main results.        
\section{The model and basic equations} 
\label{sec-model}
\begin{figure} 
\centering
\includegraphics[width=2.5in,height=2.0in]{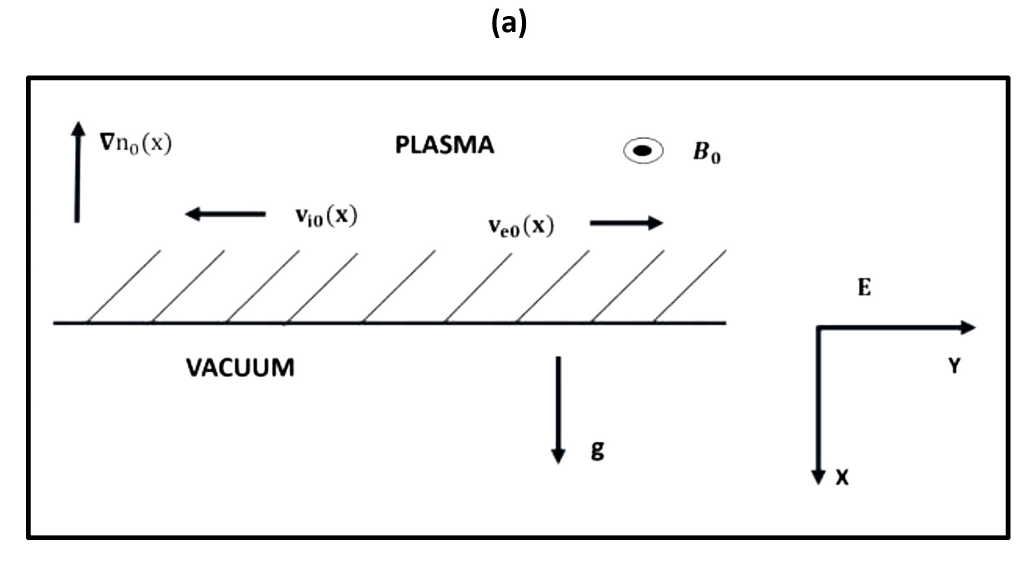}
\includegraphics[width=2.5in,height=2.0in]{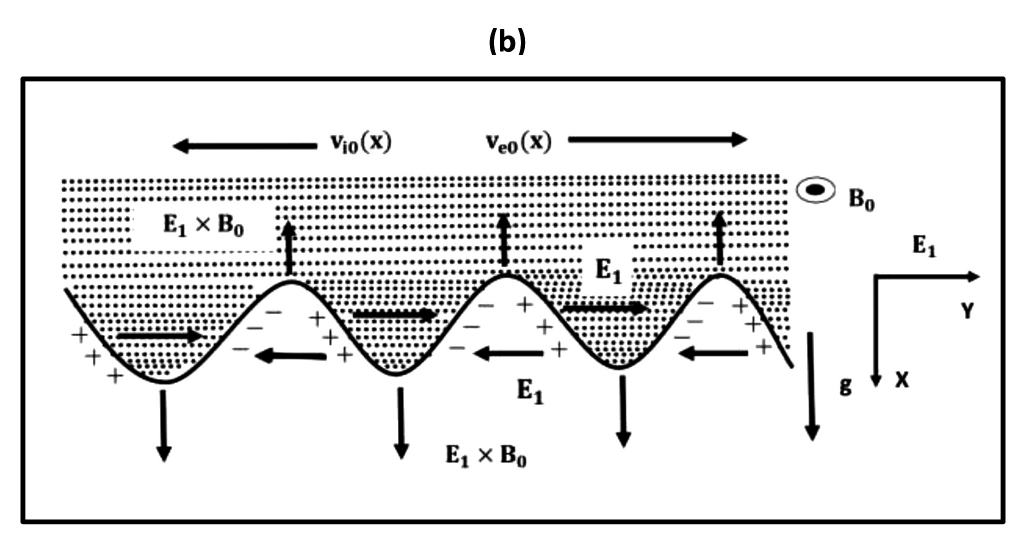}
\caption{A schematic diagram for RTI showing the equilibrium state [Subplot (a)] and the perturbed state [Subplot (b)] of an electron-ion plasma under the influences of the gravity force, the electromagnetic fields, and the density gradient. }
\label{fig:RTI}
\end{figure}

 We consider an inhomogeneous magnetized collisionless two-component electron-ion (e-i) plasma with a relativistic flow of electrons, which may be in isothermal equilibrium (nondegenerate or classical), fully degenerate (at zero temperature), or partially degenerate at finite temperature, and nondegenerate relativistic classical thermal ions. As shown in Fig. \ref{fig:RTI}, we consider a plasma boundary in the $yz$-plane immersed in an external static uniform magnetic field, ${\bm B}_0=B_0 \hat{z}$ (In low-$\beta$ plasmas, the magnetic field may be considered to be uniform) under the action of the constant gravitational field, ${\bm g}=g \hat{x}$ acting vertically downward.  We assume that the background number density gradients for electrons and ions are along the negative $x$-axis, the wave propagation vector is  ${\bm k}=(k_x,k_y,k_z)$, and the electric field is ${\bm E}=(0,E_y,0)$. At equilibrium, the uniform magnetic field acts as a light fluid to support the heavy two-component e-i plasmas and the Lorentz force balances the combined influence of the gravitational field force and the pressure gradient force. As a result, electrons and ions will drift (gravitational and diamagnetic drifts) in the opposite $y$-direction. If a perturbation in the interface develops due to random pressure fluctuations, these drift velocities of electrons and ions will cause charge separation and hence the generation of the perturbed electric field ${\bm E}$ (which changes sign as one goes from crest to trough in the perturbation).  The latter will then produce ${\bm E}\times {\bm B}_0$-drift velocities for electrons and ions [in the upward (downward) direction where the plasma layer has moved upward (downward)]. The perturbation tends to grow at the expense of the potential energy of plasma (heavier) fluids in the gravitational field, which opposes the density gradient, and when the  ${\bm E}\times {\bm B}_0$-drift velocities are properly phased. This leads to spikes in the heavier fluid penetrating lighter ones and the interface becomes unstable to form the RTI \cite{chen2012introduction}.

\par 
 As a starting point and to describe the dynamics of  relativistic compressible electron and ion fluids in quasineutral plasmas, we consider  the following dimensionless continuity and the momentum  balance equations \cite{misra2024transverse,misra2018stimulated}.
 \begin{equation}
 \frac{\partial \left(\gamma_jn_j\right)}{\partial t} +{\mathbf\nabla}\cdot (\gamma_jn_j{\bm v}_j)=0,\label{eq-2}
 \end{equation}
 \begin{equation}
  \begin{split}
\frac{{\mathrm d}}{{\mathrm {dt}}}\left(\frac{H_j \gamma_j {\bm v}_j}{n_j}\right)+\frac{\nabla P_j}{\gamma_jn_j}=\frac{m_iq_j}{m_je}\left({\bm E}+\Omega_i{\bm v}_j\times{\hat{z}}\right)+\tilde{\bm g}, \label{eq-1}
 \end{split}
 \end{equation}
 where $\mathrm{d/dt}\equiv\partial/\partial t+ \left({\bm v}_j\cdot \nabla\right)$ is the total derivative, and the quantities $n_j$, ${\bm v}_j$, and $P_j$  are, respectively,  the number density, velocity, and pressure  of $j$-th species particles [$j=i~(e)$ for ions (electrons)], normalized  by the unperturbed value $n_{j0}(0)$,  the speed of light in vacuum $c$, and $m_jc^2n_{j0}(0)$  with $m_j$ denoting the mass of $j$-th species particles. Also, the electric field  ${\bm E}$  is  normalized by $\left({m_ic^2}/{e\lambda_s} \right)$, $\Omega_i\left(=\omega_{ci}/\omega_{pi}\right)$ is the ratio between the  ion-cyclotron ($\omega_{ci}=eB_0/m_ic$) and ion plasma oscillation [$\omega_{pi}=\left(4\pi n_{i0}(0)e^2/m_i\right)^{1/2}$] frequencies, $\Omega_e=m\Omega_i$ is the normalized electron-cyclotron frequency, $m=m_i/m_e$ is the ion to electron mass ratio, and   $\tilde{\bm{g}}~\left(={\bm {g}}/{c\omega_{pi}}\right)$ is the dimensionless gravity force per unit mass of the fluid. Furthermore, $q_j=e~(-e)$ is the charge for positive ions (electrons), $H_j={\cal E}_j+P_j$ is the enthalpy per unit volume of each fluid species $j$ measured in the rest frame normalized by $m_jc^2n_{j0}(0)$ with   ${\cal E}_j$ denoting  the total energy density, $\gamma_j~\left(=1/{\sqrt{1-v_j^2}}\right)$ is the Lorentz factor for the $j$-th species particle, and  the time $t$ and  the space $x$ variables are normalized by the ion plasma period $\omega_{pi}^{-1}$ and the ion plasma skin depth $\lambda_s~(=c/\omega_{pi})$ respectively.  
\par 
Due to their heavy inertia (compared to electrons), relativistic ions are treated as classical and nondegenerate. However, depending on the plasma environments (from laboratory low-density plasmas to astrophysical degenerate dense plasmas) electrons can deviate from the thermodynamic equilibrium and obey the Fermi-Dirac statistics. In the interior of stellar compact objects such  as those of neutron stars and white dwarfs, not only electrons have  a relativistic speed but also have an arbitrary degree of degeneracy, i.e., they are either partially degenerate at finite temperature or fully degenerate at zero temperature. Thus, their pressure equations vary. In Sec. \ref{sec-linear}, we will discuss the RTI in three different cases separately, using three different equations of state for electrons, and show how the instability growth rates differ in different plasma environments. 
 \par 
 In what follows, we consider the variations of the unperturbed number density and the pressure along the $x$-axis. For adiabatic thermal ions, we write the equation of state as 
\begin{equation}
\frac{P_i}{P_{i0}(x)}=\left(\frac{n_i}{n_{i0}(x)}\right)^\Gamma,\label{eq-Pi0}
\end{equation}
where $P_{i0}(x)~\left(=\beta_i n_{i0}(x)\right)$ is the unperturbed ion pressure normalized by $n_{i0}(0)m_ic^2$ and $\beta_{i,e}=k_{\mathrm B}T_{i,e}/{m_{i,e}c^2}$ is the relativity parameter for thermal ions (electrons). We also denote $\sigma=T_i/T_e$ as the ion to electron temperature ratio. Note that while $\beta_i$ is a parameter for the cases of both classical and degenerate (partially and fully) plasmas, the parameter  $\beta_e$ appears only in classical and partially degenerate plasmas.   The normalized enthalpy for the ion species is $H_i=P_i\Gamma/(\Gamma-1)+n_i$, where the polytropic  index $\Gamma$ varies in $4/3\le\Gamma\le5/3$. In the classical limit of ion thermal motions, $\Gamma=5/3$ and $P_i\ll n_i$, while in the ultra-relativistic limit, we have $\Gamma=4/3$ and $P_i\gg n_i$. Next, we assume the unperturbed plasma number densities and the pressure inhomogeneities to vary as 
\begin{equation}
n_{j0}(x),~P_{j0}(x)\sim\exp\left(-x/L_n\right),\label{eq-nj0}
\end{equation}
 where $L_n$ is the  length scale of inhomogeneity normalized by $\lambda_s$.
\section{Rayleigh-Taylor instability} \label{sec-linear}
In this section, we study the RTI of electrostatic plane wave perturbations in  relativistic e-i magnetoplasmas by deriving a general linear dispersion relation in the low-frequency limit: $(\omega-k_yv_{j0}(x))^2 \ll \Omega_j^2;~(j=i,e)$, where $\omega~(k_y)$ is the wave frequency (wave number) of perturbations and $v_{j0}(x)$ is the drift velocity of the $j$-th species particle along the $y$-axis.  However, before deriving the dispersion relation for the perturbation, it is pertinent to consider the equilibrium plasma state.
\par 
 At equilibrium, the momentum equation for the $j$-th species particles gives
\begin{equation}
\begin{split}
&\left({\bm v}_{j0}(x)\cdot  \nabla\right)\left(\frac{H_{j0}(x)\gamma_{j0}(x) {\bm v}_{j0}(x)}{n_{j0}(x)}\right)\\
&+\frac{\nabla P_{j0}(x)}{\gamma_{j0}(x)n_{j0}(x)}=  \frac{m_iq_j}{m_je}\Omega_i\left({\bm v}_{j0}(x)\times \hat{z}\right)+ \tilde{\bm g},
\label{eq-equi}
\end{split} 
\end{equation}
from which we obtain the drift velocities for ions and electrons as \cite{chen2012introduction}
\begin{equation}\label{eq-drift-i}
\begin{split}
{\bm v}_{i0}(x)\approx -\frac{1}{\Omega_i}\left[\tilde{g}
+\frac{\beta_i }{\gamma_{i0}(0)L_n}\right] \hat{y},
\end{split}
\end{equation}
\begin{equation}\label{eq-drift-e}
\begin{split}
&{\bm v}_{e0}(x)\approx \frac{1}{\Omega_e}\left[ \tilde{g}+\frac{P_{e0}(0)}{\gamma_{i0}(0)L_n}\right] \hat{y},
\end{split} 
\end{equation}
where  $H_{e0}(x)=\alpha_e n_{e0}(x)$ and $H_{i0}(x)/n_{i0}(x)=1+\beta_i\Gamma/(\Gamma-1)$. Also,
$\gamma_{j0}(x)\approx\gamma_{j0}(0)=\left[1-v_{j0}^2(0)\right]^{-1/2}$ and
  $L_n^{-1}=-\left(1/n_{j0}(x)\right)\left(\partial n_{j0}(x)/{\partial x}\right)=-\left(1/P_{j0}(x)\right)\left(\partial P_{j0}(x)/{\partial x}\right)$. From  Eqs.  (\ref{eq-drift-i})-(\ref{eq-drift-e}), it is clear that the ions and electrons drift in opposite directions along the $y$-axis with velocities $v_{i0}(x)$ and $v_{e0}(x)$ due to the gravitational and diamagnetic drifts. Note that at equilibrium, there is no DC electric field. So, the ${\bm E}\times {\bm B}_0$-drift velocity is zero.  
  \par 
To obtain the dispersion relation for electrostatic perturbations, we split up the dependent variables into their equilibrium and perturbation parts, i.e. $n_j=n_{j0}(x)+n_{j1},~ {\bm v}_j={\bm v}_{j0}(x)+{\bm v}_{j1},~{\bm E}=0+{\bm E}_1,~ P_e=P_{e0}(x)+P_{e1}$, etc. Assuming the perturbations to vary as plane waves of the form  $\sim\exp({\mathrm i}{\bm k} \cdot{\bm r}-{\mathrm i}\omega t)$, where the wave  vector ${\bm k}$ and the wave frequency $\omega$ are normalized by $\lambda_s^{-1}$ and $\omega_{pi}$ respectively,  we obtain from Eq. \eqref{eq-1}, the following reduced equation.
\begin{equation}
\begin{aligned}
&-{\mathrm i}\left(\omega-k_yv_{i0}(x)\right)\left[\frac{\gamma_{i0}(x){\bm v}_{i0}(x)n_{i1}}{n_{i0}(x)}\left(\frac{H_{i1}}{n_{i1}}-\frac{H_{i0}(x)}{n_{i0}(x)}\right)+ \right.\\ & \left.\frac{H_{i0}(x)}{n_{i0}(x)}\left(\gamma_{i1}{\bm v}_{i0}(x)+\gamma_{i0}(x){\bm v}_{i1}\right)\right]={\bm E}_{1}+\Omega_i\left({\bm v}_{i1}\times \hat{z}\right),\label{eq-Perturbed-ion}
\end{aligned}
\end{equation}
where  $H_{i1}/n_{i1}=1+\beta_i\Gamma^2/(\Gamma-1)$. From Eq. (\ref{eq-Perturbed-ion}), the perpendicular components of the ion velocity are obtained as \cite{maryam2021rayleigh}
\begin{equation}
v_{i1x}=\frac{E_{1y}}{\Omega_i},\label{eq-vel-ion-x}
\end{equation}
\begin{equation}
v_{i1y}=-{\mathrm i}\alpha_i\frac{\left(\omega-k_yv_{i0}(x)\right)}{\Omega_i}\frac{E_{1y}}{\Omega_i},\label{eq-vel-ion-y} 
\end{equation}
where $\alpha_i\approx \gamma_{i0}(0) H_{i0}(x) /{n_{i0}(x)}$ and  $\gamma_{i1}= v_{i0}(x)v_{i1y}$.
\par 
Next, from Eq. (\ref{eq-2}), the linearized ion continuity equation gives 
\begin{equation}
\begin{split}
&\left[\omega-k_yv_{i0}(x)\right]\left[\gamma_{i0}(0)n_{i1}+\gamma_{i1}n_{i0}(x)\right]-\gamma_{i0}(0)k_y n_{i0}(x)v_{i1y}\\ & -\gamma_{i0}(0)n_{i0}(x)\left(k_x+\frac{{\mathrm i}}{L_n}\right)v_{i1x}=0.\label{eq-perturbed-cont}
\end{split}
\end{equation}

 Substituting Eqs. (\ref{eq-vel-ion-x}) and (\ref{eq-vel-ion-y}) into Eq. (\ref{eq-perturbed-cont}),  we obtain 
\begin{equation}
\frac{E_{1y}}{\Omega_i}=\frac{{\mathrm i}n_{i1}}{n_{i0}(x)}\left(\frac{\omega-k_yv_{i0}(x)}{\alpha_i k_y\left(\omega-k_yv_{i0}(x)\right)/\Omega_i-1/L_n+{\mathrm i}k_x}\right).\label{eq-con-mom-ion}
\end{equation}
This expression for the electric field perturbation agrees with Ref. \cite{maryam2021rayleigh} after one has fixed the normalization for the physical quantities.
Typically, the electron drift velocity is smaller than ions. So, in the expressions where both appear, we can safely ignore the electron-drift velocity compared to the ions. Thus, assuming the smallness of $v_{e0}(x)$ so that $\gamma_{e0}(0)\sim 1$  in the limit $m_e/m_i\rightarrow 0$, we get from Eq. \eqref{eq-1} the following expressions for the electron velocity (perturbed) components.
\begin{equation}
\begin{aligned}
&v_{e1x}=\frac{E_{1y}}{\Omega_i}+\frac{{\mathrm i}n_{e1}}{n_{e0}(x)\Omega_e}\left[k_yb+\frac{\omega\alpha_e}{\Omega_e}\left(\frac{P_{e0}(0)}{L_n}+{\mathrm i}k_xb\right)\right],\label{eq-vel-x-elec}
\end{aligned}
\end{equation}

\begin{equation}
v_{e1y}=-\frac{ n_{e1}}{ n_{e0}(x)\Omega_e}\left(\frac{P_{e0}(0)}{L_n}+{\mathrm i}bk_x\right),\label{eq-vel-y-elec}
\end{equation}
\begin{equation}
v_{e1z}=\frac{k_zP_{e1}}{\omega n_{e0}(x)\Omega_e}, \label{eq-vel-z-elec}
\end{equation}
where  $b=P_{e1}/n_{e1}$. Substituting  Eqs. (\ref{eq-vel-x-elec})-(\ref{eq-vel-z-elec}) into the linearized electron continuity equation [to be obtained from Eq. (\ref{eq-2}) for electrons, i.e., for $j=e$], we obtain
  \begin{equation}
\begin{split}
&\frac{E_{1y}}{\Omega_i}=-\frac{{\mathrm i}n_{e1}}{n_{e0}(x)} \left[\omega L_n\left(\frac{\alpha_e P_{e0}(0)}{\Omega_e^2 L_n^2}+\frac{1}{\left(1-{\mathrm i} L_nk_x\right)}\right) \right. \\ 
&\left. 
+\frac{k_y}{\Omega_e}\left(\frac{P_{e0}(0)}{\left(1-{\mathrm i} L_nk_x\right)}+b\right)+\frac{{\mathrm i}b\alpha_e k_x\omega}{\Omega_e^2}\right. \\ 
&\left. +\frac{bL_n}{\left(1-{\mathrm i} L_nk_x\right)}\left(\frac{{\mathrm i}k_x k_y}{\Omega_e}-\frac{k_z^2}{\omega \alpha_e}\right)\right].\label{eq-con-mom-ele}
\end{split}
\end{equation}

  Next, eliminating $E_{1y}/\Omega_i$ from Eqs.  (\ref{eq-con-mom-ion}) and (\ref{eq-con-mom-ele}),  and using the quasi-neutrality conditions: $n_{e1}\simeq n_{i1}$ and  $n_{i0}(x)=n_{e0}(x)= n_0(x)$ for the perturbed and unperturbed states,
 we obtain the following three-dimensional dispersion relation for  electrostatic perturbations in an electron-ion magnetoplasma under the influence of gravity.
 
\begin{equation}
A_1\omega^3+B_1\omega^2+C_1\omega+D_1=0, \label{eq-disper-22}
\end{equation}
where the coefficients $A_1$, $B_1$, $C_1$, and $D_1$ are given by

 \begin{widetext}
\begin{equation}
A_1=\left(\frac{1}{\left(1-{\mathrm i}L_nk_x\right)}+\frac{\alpha_eP_{e0}(0)}{\Omega_e^2L_n^2}+\frac{{\mathrm i}b\alpha_ek_x}{\Omega_e^2L_n}\right),\label{exp-A_1}
\end{equation}
\begin{equation}
\begin{aligned}
B_1=&k_y\left[\left(\frac{\tilde g}{\Omega_i}+\frac{\beta_i}{L_n\Omega_i\gamma_{i0}(0)}\right)A_1-\frac{\left(1-{\mathrm i}L_nk_x\right)\alpha_eP_{e0}(0)\Omega_i}{\alpha_ik_y^2L_n^3\Omega_e^2}+\frac{1}{L_n\Omega_e}\left(\frac{P_{e0}(0)}{\left(1-{\mathrm i}L_nk_x\right)}+b\right)\right.\\&\left.+\frac{{\mathrm i}bk_x}{\left(1-{\mathrm i}L_nk_x\right)\Omega_e}\left(1-\frac{\left(1-{\mathrm i}L_nk_x\right)^2\alpha_e\Omega_i}{\alpha_ik_y^2L_n^2\Omega_e}\right)\right],
\end{aligned}\label{exp-B_1}
\end{equation}
\begin{equation}
\begin{aligned}
C_1= &\frac{\tilde g}{\alpha_iL_n}\left[\Big\lbrace1+\frac{\alpha_ik_y^2}{\Omega_e\Omega_i}\left(\frac{P_{e0}(0)}{\left(1-{\mathrm i}L_nk_x\right)}+b\right)+\frac{{\mathrm i}bk_xk_y^2\alpha_iL_n}{\left(1-{\mathrm i}L_nk_x\right)\Omega_e\Omega_i}\Big\rbrace\left(1+\frac{\beta_i}{L_n\tilde g\gamma_{i0}(0)}\right)\right.\\&\left.-\frac{\left(1-{\mathrm i}L_nk_x\right)\Omega_i}{\tilde{g}L_n\Omega_e}\left(\frac{P_{e0}(0)}{\left(1-{\mathrm i}L_nk_x\right)}+b\right)-\frac{b\alpha_iL_nk_z^2}{\left(1-{\mathrm i}L_nk_x\right)\alpha_e\tilde g}-\frac{{\mathrm i}bk_x\Omega_i}{\Omega_e\tilde g}\right],
\end{aligned}\label{exp-C_1}
\end{equation}
\begin{equation}
D_1=\frac{b\Omega_ik_z^2}{\alpha_ik_yL_n\alpha_e}\left[1-\frac{\alpha_ik_y^2L_n}{\left(1-{\mathrm i}L_nk_x\right)\Omega_i^2}\left(\tilde g+\frac{\beta_i}{\gamma_{i0}(0)L_n}\right)\right].
\end{equation}

\end{widetext}
Without loss of generality, by restricting the wave propagation in the $yz$-plane and assuming the wave perturbations along the $x$-axis to be small compared to the $y$- and $z$-axes and the length scale of density inhomogeneity to be much smaller than the perturbed wavelength along the $x$-axis, we obtain from Eq. \eqref{eq-disper-22} the following reduced dispersion equation with real coefficients.  
\begin{equation}
\begin{split}
A\omega^3&+B\omega^2+\left(C-\frac{bk_z^2}{\alpha_e}\right)\omega\\
&+\frac{bk_yk_z^2}{\alpha_e\Omega_i}\left[\frac{\Omega_i^2}{\alpha_ik_y^2L_n}-\left(\tilde g +\frac{\beta_i}{\gamma_{i0}(0)L_n}\right)\right], \label{eq-disper-33}
\end{split}
\end{equation}
where the coefficients $A$, $B$, and $C$ are given by
\begin{equation}
A=\left(1+\frac{\alpha_eP_{e0}(0)}{\Omega_e^2L_n^2}\right),\label{exp-A}
\end{equation}
\begin{equation}
\begin{aligned}
B=&k_y\left[\left(\frac{\tilde g}{\Omega_i}+\frac{\beta_i}{L_n\Omega_i\gamma_{i0}(0)}\right)\left(1+\frac{\alpha_eP_{e0}(0)}{\Omega_e^2L_n^2}\right)\right.\\&\left.-\frac{\alpha_eP_{e0}(0)\Omega_i}{\alpha_ik_y^2L_n^3\Omega_e^2}+\frac{1}{L_n\Omega_e}\left(P_{e0}(0)+b\right)\right],
\end{aligned}\label{exp-B}
\end{equation}
\begin{equation}
\begin{aligned}
C= &\frac{\tilde g}{\alpha_iL_n}\left[\Big\lbrace1+\frac{\alpha_ik_y^2}{\Omega_e\Omega_i}\left(P_{e0}(0)+b\right)\Big\rbrace\left(1+\frac{\beta_i}{L_n\tilde g\gamma_{i0}(0)}\right)\right.\\&\left.-\frac{\Omega_i}{\tilde{g}L_n\Omega_e}\left(P_{e0}(0)+b\right)\right].
\end{aligned}\label{exp-C}
\end{equation}  
 
\par 
Inspecting on the coefficients $A,~B$ , and $C$ of the cubic equation \eqref{eq-disper-33}, we observe that   
$A,~B,~C>0$, and the constant term is negative for the parameters relevant for classical and degenerate plasmas, to be discussed shortly. Thus, by Descarte's rule of sign, Eq. \eqref{eq-disper-33} has either one real positive and two real negative roots or one positive real root and a pair of complex conjugate roots. Furthermore, the coefficients $A$, $B$, and $C$ are responsible for the existence of either negative real roots or complex conjugate roots. Since we are interested in the instability (i.e., imaginary parts of the complex roots), not the propagating mode (associated with the real root or real part of the complex root), it is sufficient to disregard the constant term in Eq.    \eqref{eq-disper-33} (since the leading term in $\omega$ gives the desired root), which can be possible for further restricting the wave propagation along the $y$-axis (i.e., when $k_x,~k_z=0$) and similar conditions on the wave perturbations and the density inhomogeneity scale imposed on Eq. \eqref{eq-disper-22} to get Eq. \eqref{eq-disper-33}. Thus, from Eq. \eqref{eq-disper-33}, we obtain  

\begin{equation}
A\omega^2+B\omega+C=0. \label{eq-disper-2}
\end{equation}

We note that Eq. (\ref{eq-disper-2}) agrees with the dispersion relation obtained in \cite{maryam2021rayleigh} after one disregards the radiation pressures therein and fixes the normalization for the variables.  Equation (\ref{eq-disper-2}) gives two solutions for the wave frequency:
\begin{equation}
\omega=-\frac{B}{2A}\pm\frac{\left(B^2-4AC\right)^{1/2}}{2A}.\label{exp-omega}
\end{equation}
Thus, electrostatic perturbations are unstable and the RTI sets in if the wave frequency becomes complex, i.e., if
 \begin{equation}
B^2-4AC<0.\label{con-RTI-1}
\end{equation}
Since we are interested in the instability rather than damping, so taking the plus sign in Eq. (\ref{exp-omega}) before the radical sign (Since $A>0$) and assuming $\omega =\omega_r+{\mathrm i}\omega_i$, we obtain
\begin{equation}
\omega_r=-\frac{B}{2A},\label{exp-real-omega}
\end{equation}
\begin{equation}
\omega_i=\frac{\left(4AC-B^2\right)^{1/2}}{2A}.\label{exp-complex-omega}
\end{equation}
Note that the coefficients $A$, $B$, and $C$, given by  Eqs. (\ref{exp-A}) and (\ref{exp-C}), are significantly modified by the electron and ion pressures and the relativistic dynamics of electrons and ions. In particular, in the absence of electron pressure (i.e., $P_{e0}(0)=0$ and $b=0$ ), Eqs. (\ref{exp-real-omega}) and (\ref{exp-complex-omega}) reduce to
\begin{equation}
\omega_r=\frac{1}{2}\frac{k_y\tilde{g}}{\Omega_i},\label{exp-real-omega-P0}
\end{equation}
\begin{equation}
\omega_i=\sqrt{\tilde{g}\left(\frac{1}{\alpha_iL_n}-\frac{4k_y^2\tilde{g}}{\Omega_i^2}\right)}.\label{exp-complex-omega-P0}
\end{equation}
From Eqs. (\ref{exp-real-omega-P0}) and (\ref{exp-complex-omega-P0}), it follows that in absence of the electron pressure, the real mode propagates with a constant phase velocity (i.e., dispersionless), which is reduced by the effects of strong magnetic field, and the RTI occurs in the domain:   $0<k_y<k_c\equiv\sqrt{{\Omega_i^2/{4\alpha_i L_n \tilde{g}}}}$. Thus, as the wave number increases but remains smaller than the critical value $k_c$, the growth rate tends to decrease after reaching a maximum value at $k_y=0$. In the limit of long-wavelength perturbations, i.e., $k_y\to 0$, the growth rate becomes constant, i.e., $\omega_i\sim \sqrt{\tilde{g}/{\alpha_iL_n}}$, and by disregarding the Lorentz (relativistic) factor, one can recover the known result in the literature \cite{chen2012introduction} $\omega_i\sim \sqrt{\tilde{g}/{L_n}}$. Note that the instability domain of $k_y$ expands and the instability growth rate increases either by increasing the magnetic field strength or by reducing the length scale of density inhomogeneity. Physically, as the magnetic field increases, the ${\bm E}\times {\bm B}_0$-drift velocity increases, which results into the generation of more spikes in plasmas (heavier fluids) penetrating the lighter fluids with enhanced amplitudes.
 However, in a general situation, the instability domain may be in the form   $k_{c^-}<k_y<k_{c^+}$ and the instability growth curve can be in a parabolic form, where the extremities $k_{c^-}$ and $k_{c^+}$ are to be determined. Since explicit expressions for $k_{c^+}$ and $k_{c^-}$ are difficult  to obtain, one can find their values numerically with the variations of different plsma parameters, namely $\beta_e,~\beta_i,~\Omega_e,~L_n$, etc. In Secs. \ref{sec-classical}-\ref{sec-PDC}, we will study the instability growth rates in different plasma environments, especially when electrons are (i) classical (nondegenerate) or in isothermal equilibrium, (ii) fully degenerate (forming a zero-temperature Fermi gas), and (iii) partially degenerate or degenerate at finite temperature.  
\subsection{Classical or nondegenerate electrons}\label{sec-classical}
We begin our study on RTI in a classical relativistic magnetoplasma under gravity in which we consider an adiabatic ion pressure but an isothermal equation of state for electrons. Such plasmas are relevant in laboratory and space plasmas where electron thermal energy can vary from nonrelativistic $(\beta_e\ll1)$ to ultra-relativistic $(\beta_e\gg1)$ regimes in comparison with the electron rest mass energy. The normalized isothermal pressure and the enthalpy for electrons are  given by \cite{berezhiani1995large}
\begin{equation}
P_{e}=\beta_en_e,~H_e=n_eG_e,\label{exp-pre-cla}
\end{equation}
where $G_e$ is the effective mass factor such that $G_e(z_{cl})=K_3(z_{cl})/{K_2(z_{cl})}$ with $K_3(z_{cl})$ and $K_2(z_{cl})$, respectively, denoting the MacDonald functions of the second and third orders, and $z_{cl}=1/{\beta_e}$.  
In particular, in the non-relativistic or weakly relativistic thermal motion of electrons $(\beta_e\ll 1)$,  we have $G_e\sim (1+{5\beta_e/2})$ (i.e., $G_e>1$), which in the cold plasma limit ($T_e=0$) reduces to $G_e\sim 1$. In the opposite limit, i.e., the limit of high-temperature or ultra-relativistic motion $(\beta_e\gg 1)$, we have $G_e\sim4\beta_e$ i.e. $ G_e\gg1$. At equilibrium, we have   $H_{e0}(x)=\alpha_en_{e0}(x)~(\sim G_{e0}n_{e0}(x))$, where
\begin{equation} 
\alpha_e=
\begin{cases}
\begin{split}
 & \left(1+\frac{5}{2}\beta_e\right) , & \text{for $\beta_e\ll 1$},\\ \\
 &4\beta_e,& \text{for $\beta_e\gg 1$}.
\end{split}
\end{cases}
\end{equation}  
Also, we have the relation for the dimensionless parameter $\beta_e$: $P_{e0}(0)=P_{e1}/n_{e1}=\beta_e$. It follows that the key parameter associated with the electron thermal motion is $\beta_e$, which strongly influences the instability growth rate. We numerically analyze the instability growth rates in two different cases, namely when $\beta_e<1$ and $\beta_e>1$ as follows:
 \subsubsection*{Case I: $\beta_e<1$.}
\begin{figure} 
\centering
\includegraphics[width=3.7in,height=2.5in]{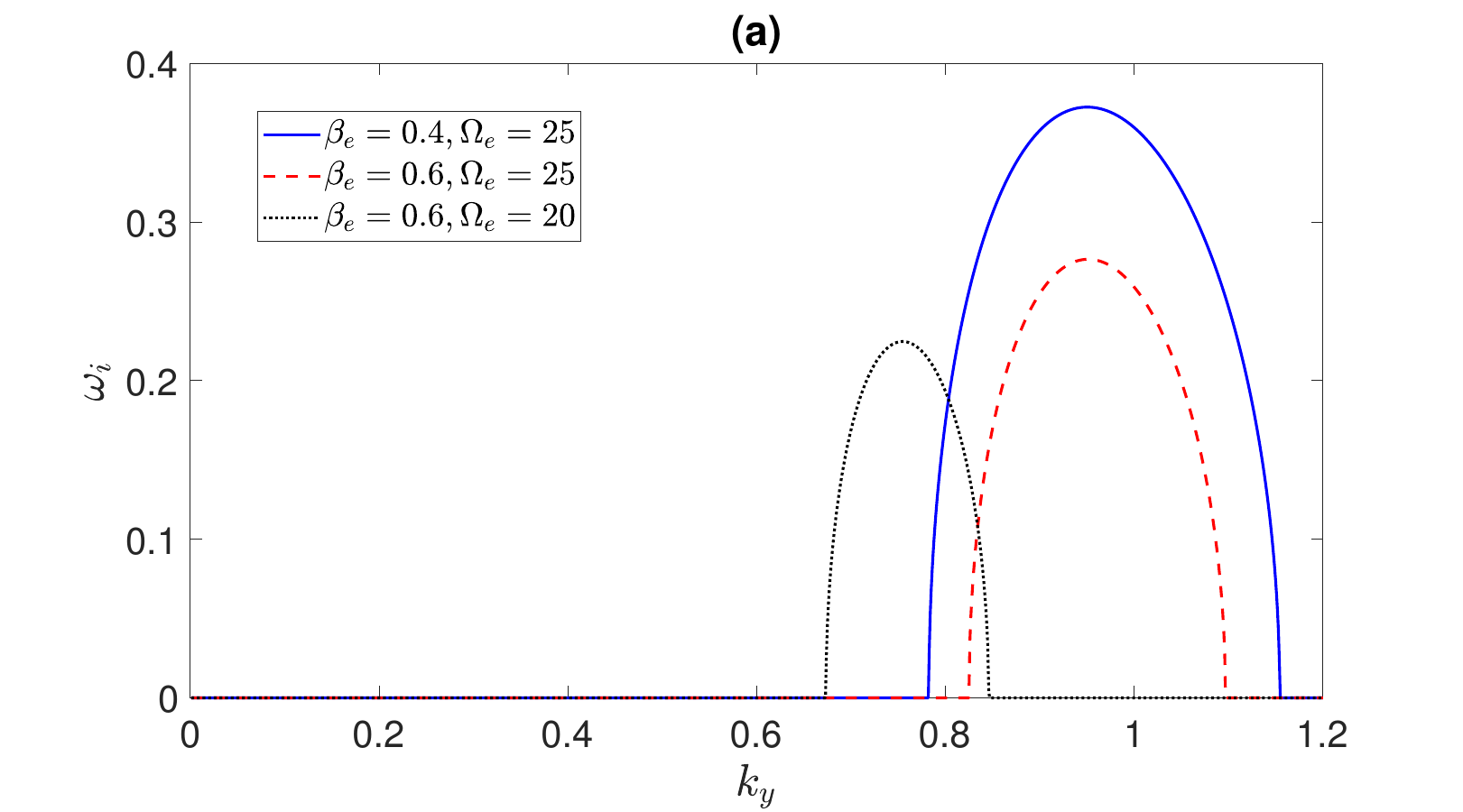}
\includegraphics[width=3.7in,height=2.5in]{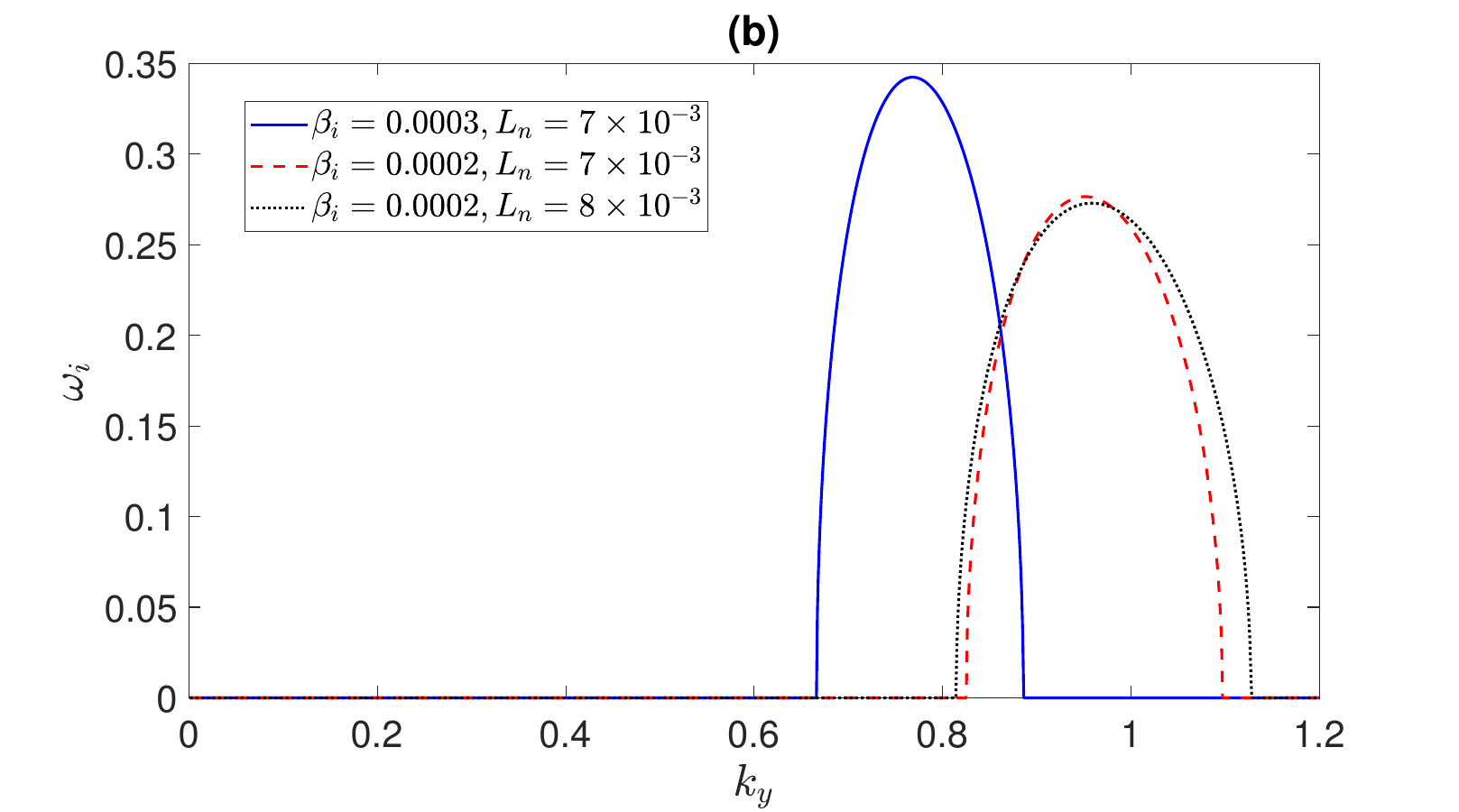}
\caption{Classical plasmas with $\beta_e<1$. The profiles of the growth rates are shown with the variations of different plasma parameters as in the legends with $m=1900$, $\Gamma=1.5$ and $\tilde{g}=2.53\times 10^{-16}$. The other fixed parameter values for subplots (a) and (b), respectively, are  ($L_n=7\times10^{-3}$ and $\beta_i=0.0002$)  and ($\beta_e=0.6$ and $\Omega_e=25$). }
\label{fig:cl_be_le_1_1}
\end{figure}

 In this case, we choose the typical parametric values relevant for laboratory plasmas \cite{borthakur2018} as $n_0(0)\simeq 10^{15}-10^{22}~ {\rm{cm}^{-3}},~ T_i\simeq3.26 \times 10^6-3.26\times 10^{10}~ {\rm K},~ T_e\simeq 5.93 \times 10^8-4.74\times 10^9~ {\rm K},~ B_0\simeq 9.47\times 10^4-2.99\times10^8~ {\rm G},~g\simeq 10^2-10^8 ~{\rm cm~ s^{-2}}$, and $L_n \simeq 6.83\times10^{-7}-7.2\times10^{-2}~ {\rm{cm}} $. Here, the electron thermal energy is close to but less than the electron rest mass energy, i.e., $\beta_e<1$, and plot the instability growth rate $\omega_i~(0\lesssim\omega_i<1)$  against the wave number $k_y~(0<k_y< 1.2)$ for different values of  $\beta_e,~\beta_i,~\Omega_e$ and $~L_n $,  and fixed values of $m, ~\Gamma$ and $\tilde{g}$ as shown in Fig.  \ref{fig:cl_be_le_1_1}.  The growth rate is reduced as the value of $\beta_e$ is increased or a value of any of $\beta_i$ and $\Omega_e$ gets reduced with cut-offs at lower values of $k_y~(<1.2)$. The instability domain, however, contracts with decreasing values of $\beta_i$  and $\Omega_e$ but increasing values of  $\beta_e$.
 Physically, as the electron's thermal energy increases, its drift velocity $v_{e0}(x)$ increases [since $P_{e0}(0)=\beta_e$] but the ion-drift velocity remains almost unchanged since $\beta_i=\beta_e\sigma/m\ll\beta_e$. As a result, electrons get separated from ions (as their drift velocity $v_{i0}(x)$ remains unchanged at this stage) to build up a perturbed electric field with lower intensity, leading to a lower value of the ${\bm E}\times {\bm B}_0$-drift velocity. The latter may not get adequately phased to enhance the spikes (\textit{cf}. Fig.\ref{fig:RTI}) and hence the instability growth gets reduced [See the solid and dashed lines of the subplot (a)]. However, when $\beta_i$ is increased, both the drift velocities of electrons and ions increase. So, a strong perturbed electric field is produced to enhance the ${\bm E}\times {\bm B}_0$-drift velocity and hence the instability growth rate [See the solid and dashed lines of subplot (b)].
Similarly, increasing the magnetic field strength enhances the ${\bm E}\times {\bm B}_0$-drift velocity and hence an increase in the instability growth rate (See the dashed and dotted lines of the subplot (a)).  By the similar reasons as stated above, since the length scale of inhomogeneity reduces the electron and ion-drift velocities, the growth rate is decreased with increasing values of $(L_n)$ [See subplot (b)].  
 In particular, for $\beta_i\sim 0.0002,~\beta_e\sim 0.6,~ \Omega_e\sim 25$, and $ L_n\sim7\times10^{-3}$, the maximum growth rate is observed to be $\omega^{\max}_{i}=0.37$ at $k_y=0.95$. 
 \subsubsection*{Case II: $\beta_e>1$.}
\begin{figure} 
\centering
\includegraphics[width=3.7in,height=2.5in]{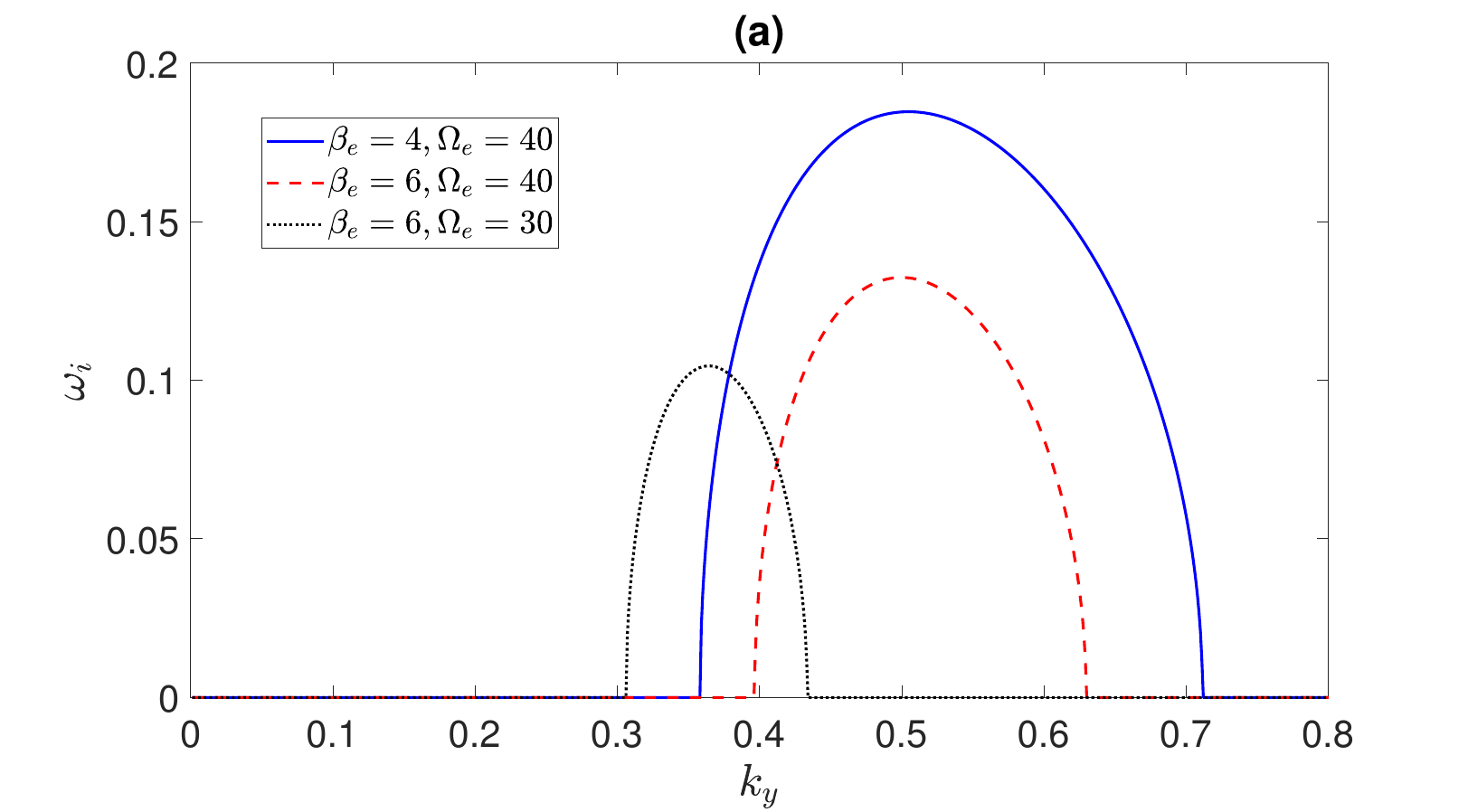}
\includegraphics[width=3.7in,height=2.5in]{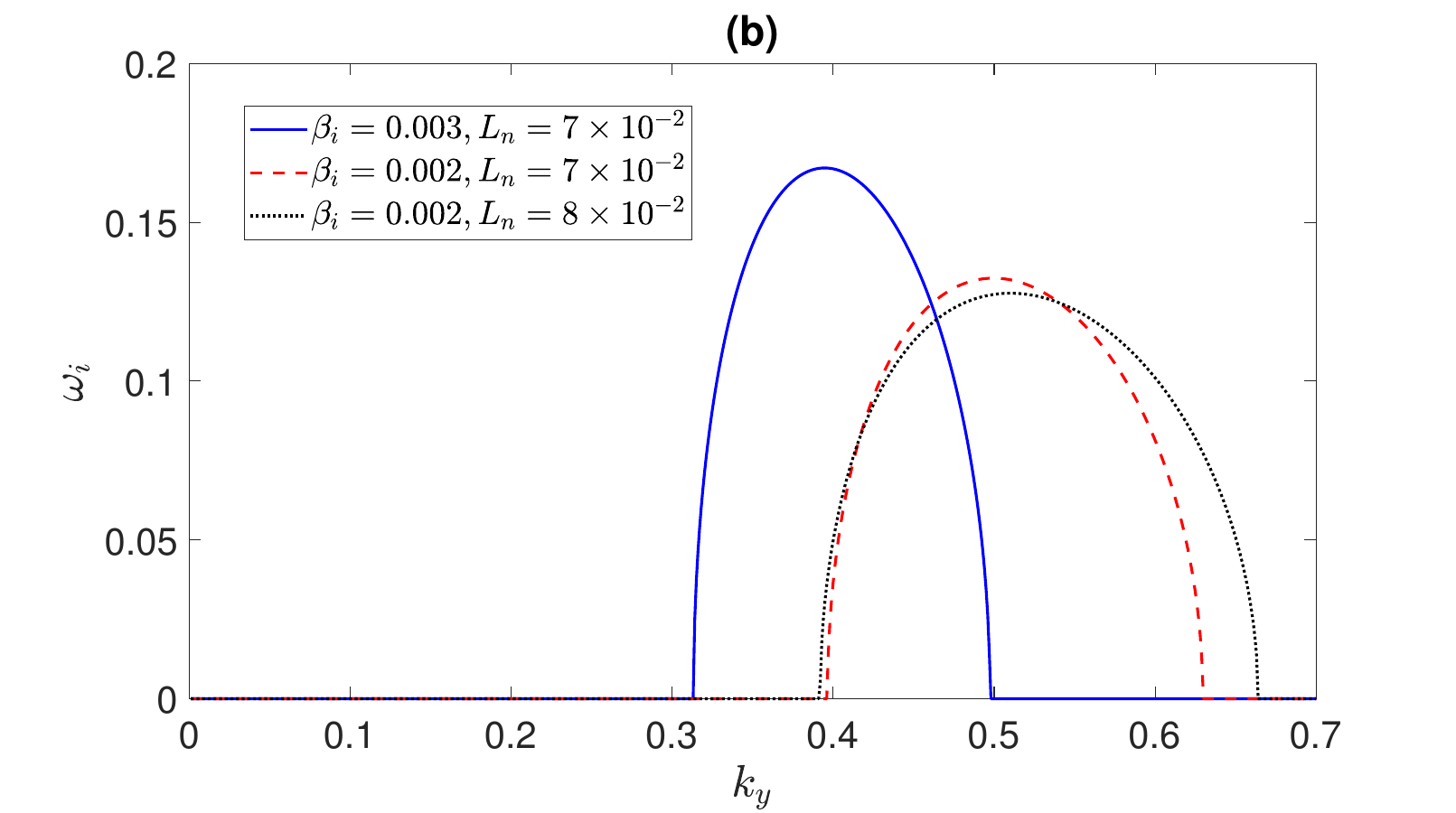}
\caption{Classical plasmas with $\beta_e>1$. The profiles of the growth rates are shown with the variations of different plasma parameters as in the legends with $m=1900$, $\Gamma=1.5$, and $\tilde{g}=2.53\times 10^{-16}$. The other fixed parameter values for subplots (a) and (b), respectively, are  ($L_n=7\times10^{-2}$ and $\beta_i=0.002$)  and ($~\beta_e=6$ and $\Omega_e=40$).}
\label{fig:cl_be_gt_1_1}
\end{figure}

  In the case of larger thermal energy than the rest mass energy of electrons, we choose the typical parametric values relevant for laboratory plasmas \cite{borthakur2018} as  $n_0(0)\simeq 10^{15}-10^{22}~ {\rm cm^{-3}},~ T_i\simeq4.35 \times 10^7-4.35\times 10^9~ {\rm K},~ T_e\simeq 1.18 \times 10^{10}-4.74\times 10^{10}~ {\rm K},~ B_0\simeq 9.47\times 10^4-2.99\times10^8~ {\rm G},~g\simeq 10^2-10^8 ~{\rm cm~ s^{-2}}$, and $L_n \simeq 6.83\times10^{-6}-7.2\times10^{-1}~{\rm  cm} $. A reduction in the growth rate, similar to Case I, is also noted. However, the reduction becomes significant with increasing values of $\beta_e$, and decreasing values of the magnetic field strength $(\Omega_e)$ and the ion thermal energy $(\beta_i)$. Not only are the growth rates reduced like in Case I,  the instability domains also significantly shrink with increasing values of $\beta_e$ and $\beta_i$ but decreasing values of $L_n$.  The results are displayed in Fig. \ref{fig:cl_be_gt_1_1}. Also, similar to Case I, the instability domains shift towards lower (or higher) values of $k_y$ as the magnetic field strength decreases (or the length scale, $L_n$ increases or the ion thermal energy decreases relative to the rest mass). Thus, it follows that although the qualitative features of the instability growth rates are similar, the instability growth rate is found to be higher in the case of $\beta_e<1$ compared to $\beta_e>1$ and the instability domains  significantly differ in the two regimes of electron thermal energy.   In particular, for $\beta_i\sim 0.0002,~\beta_e=4,~ \Omega_e\sim 40$, and $L_n\sim7\times10^{-2}$, the maximum growth rate is found to be $\omega^{\max}_{i}=0.18 $ at $k_y=0.51$.
\subsection{Fully degenerate electrons}\label{sec-FDC}
\begin{figure} 
\centering
\includegraphics[width=3.7in,height=2.5in]{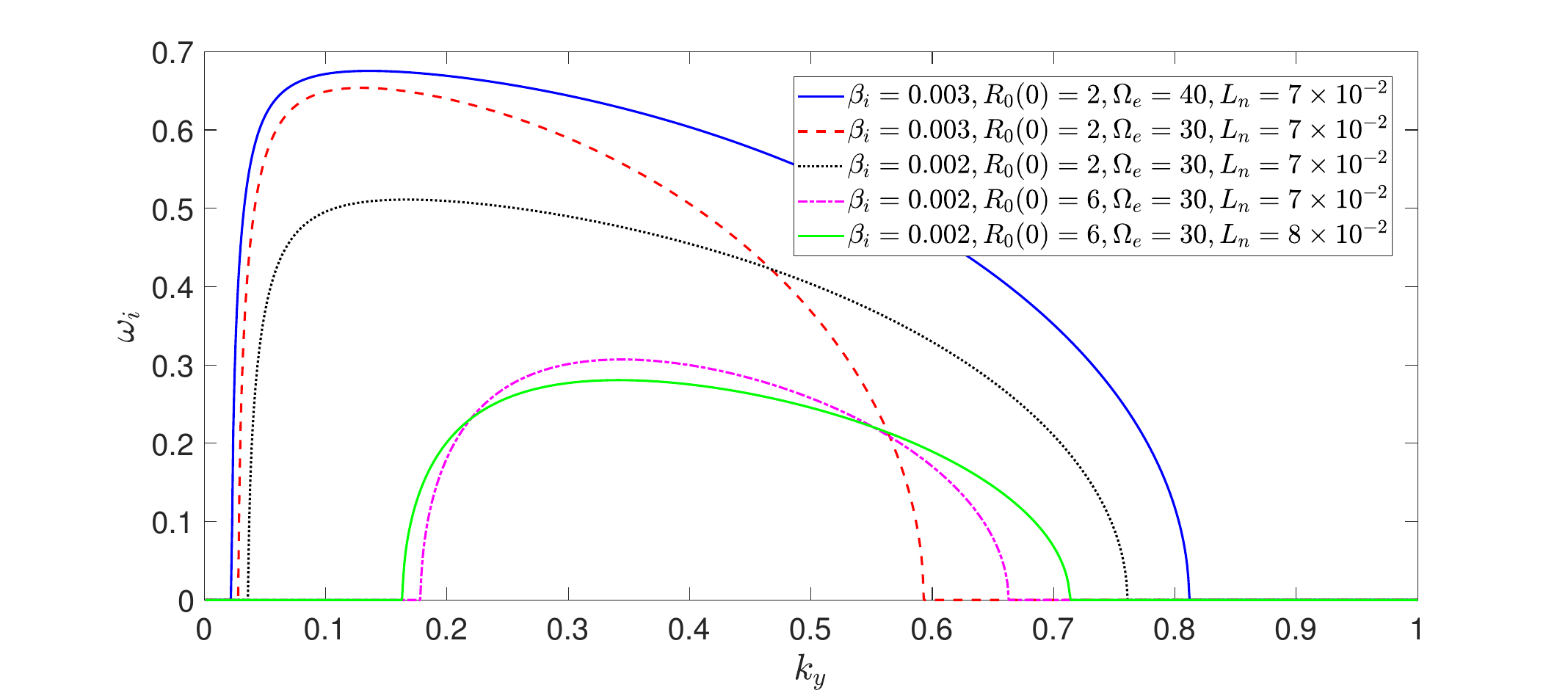}
\caption{Fully degenerate plasmas: The profiles of the growth rates are shown with the variations of different plasma parameters as in the legends with $m=1900$, $\Gamma=1.5$, and $\tilde{g}=2.80\times 10^{-21}$.}
\label{fig:fig_full_be_le_1}
\end{figure}

We consider the case in which electrons form a Fermi degenerate gas at zero temperature ($T_e=0$ K). In astrophysical environments, such as those in the core of massive stars like white dwarfs  \cite{peterson2021effects,saumon20221current}, ions are typically nondegenerate, i.e., classical ($T_i\gg T_{Fi}$) due to their inertia and electrons can be fully degenerate ($T_e\ll T_{\rm{Fe}}$) providing the degeneracy pressure to prevent the collapse of the stars. Here, $T_{Fj}$ denotes the Fermi temperature for $j$-th species particles. The Fermi-Dirac pressure law  for degenerate electrons (at $T=0$ K) gives \cite{chandrasekhar1935the,misra2018stimulated}
\begin{equation}
P_{F}=\frac{m_e^3c^3}{8\pi^2\hbar^3n_{e0}(0)}\left[R\left(1+R^2\right)^{1/2}\left(\frac{2R^2}{3}-1\right)+\sinh^{-1}R\right],\label{exp-pre-FD}
\end{equation}
where $R=R_0(0)n_e^{1/3}$ is the Fermi momentum normalized by $m_ec$ with $R_0(0)=(\hbar/m_ec)\left(3\pi^2n_{e0}(0)\right)^{1/3}$ denoting the degeneracy parameter and $\hbar=h/{2\pi}$ the reduced Plank's constant.  The normalized enthalpy  of electrons is   $H_e=n_e \sqrt{1+R^2}$. Note that, the electron degeneracy can be highly relativistic if $R\gg 1$ and non relativistic if $R\ll  1$.
In fully degenerate e-i plasmas, the Fermi-energy ($k_{\mathrm B}T_{\rm{Fe}}$) is much higher than the thermal energy ($k_{\mathrm B}T_e$) of electrons i.e., $k_{\mathrm B}T_{\rm{Fe}}\gg k_{\mathrm B}T_e $ and $R_0(0)$ measures the degree of degeneracy of unperturbed electrons. The general dispersion relation (\ref{eq-disper-2}) can now be studied numerically for fully degenerate e-i plasmas. The results are displayed in Fig. \ref{fig:fig_full_be_le_1}. In this case, the expressions for $P_{e0}(0)$  and $b$ are given by
\begin{equation}
\begin{aligned}
P_{e0}(0)=\frac{3}{8R_0^3(0)}&\left[R_0(0)\left(1+R_0^2(0)\right)^{1/2}\left(\frac{2R_0^2(0)}{3}-1\right)\right.\\ &\left.+\sinh^{-1}R_0(0)\right],
\end{aligned}\label{exp-pre-equ-FD}
\end{equation}
\begin{equation}
b=\frac{1}{3} R_0^2(0)\left[1+R_0^2(0)\right]^{-1/2}.\label{exp-b-FD}
\end{equation}
We choose the typical parametric values relevant for the environments of magnetized white dwarfs   \cite{peterson2021effects,saumon20221current} as: $n_0(0)\simeq 10^{30}-10^{32}~{\rm cm^{-3}},~T_i=3.26 \times 10^6-3.26\times 10^{9}~ {\rm K},~B_0\simeq 7.48\times 10^{10}-2.99\times 10^{13} ~{\rm G},~g\simeq 10^8 ~{\rm cm~ s^{-2}}$, and $L_n \simeq 1.13\times10^{-10}-1.59\times10^{-7} {\rm ~cm} $.  
From Fig. \ref{fig:fig_full_be_le_1}, we note that the qualitative features of the instability growth rate in fully degenerate plasmas are quite distinctive compared to the classical nondegenerate plasmas (\textit{cf}. Figs. \ref{fig:cl_be_le_1_1} and \ref{fig:cl_be_gt_1_1}). We also find that similar to Figs. \ref{fig:cl_be_le_1_1} and \ref{fig:cl_be_gt_1_1}, the instability growth rate is reduced but the instability domain expands due to reduction of the ion thermal energy compared to the rest mass energy (See the dashed and dotted lines). Furthermore, while the magnetic field strength enhances the growth rate and the instability domain (See the blue solid line and dashed lines), the degeneracy parameter, in contrast, reduces both of them (See the dotted and dash-dotted lines). 
The latter is  in agreement with the investigation \cite{bychkov2008rayleigh} where the authors have shown that the quantum pressure can weaken the RTI in quantum plasmas. 
However, the growth rate can be reduced but the instability domain can be expanded by increasing the inhomogeneity scale size (See the dash-dotted and green solid lines). In particular, for $\beta_i\sim 0.003, ~R_0(0)\sim 2,~ \Omega_e\sim 40$, and $L_n\sim7\times10^{-2}$, the maximum growth rate is found to be $\omega^{\max}_{i}\approx0.676$ at $k_y=0.14$. Thus, in comparison with the classical results, it may be concluded that the fully degenerate Fermi gas with higher relativistic degeneracy tends to weaken the RTI.
 
\subsection{Partially degenerate electrons}\label{sec-PDC}
In many astrophysical situations (e.g., the environments of white dwarfs, neutron stars, and in gas giants like Jupiter) or laser produced plasmas, the strict condition $T_e\ll T_{\rm{Fe}}$  for full degeneracy of electrons may not always be fulfilled, i.e., there may not be any strict upper limit for the energy levels. Thus, one can reasonably assume either $T_e<T_{\rm{Fe}}$  or $T_e>T_{\rm{Fe}}$. We, however, consider the regime where the electron Fermi energy is larger than the thermal energy, i.e., $T_e<T_{\rm{Fe}}$, the dimensionless electron chemical energy $\xi=\mu/k_{\mathrm B}T_e$ is positive and finite, and the electron thermal  and the rest mass energies do not differ significantly, i.e., $\beta_e\equiv k_{\mathrm B}T_e/m_ec^2\sim 1$ (implying that either $\beta_e<1$ or $\beta_e>1$). Thus, for partially degenerate electrons or electrons with arbitrary degeneracy, we 
consider the following expression for the electron number density \cite{boshkayev2016equilibrium,shah2010effect}.
\begin{equation}\label{eq-nj}
n_e=\int fd{p_e}=\frac{2}{(2\pi \hbar)^3} \int_{0}^{\infty} \frac{4\pi p_e^2}{\exp  \left(\frac{ E(p_e)- \mu}{k_{\mathrm B}T_e}\right)+1} dp_e,
\end{equation}  
where $\hbar$ is the Planck's constant divided by $2\pi$ and $\mu$ is the  chemical potential energy for electrons without the rest mass energy.  Also, $E(p_e)=\sqrt{c^2p_e^2+m_e^2c^4}$  is the relativistic  energy, and $p_e$ and $m_e$ are, respectively, the relativistic momentum and the rest mass of electrons.
\par
Equation \eqref{eq-nj} can be put into the  following alternative form  \cite{boshkayev2016equilibrium,timmes1999accuracy} 
\begin{equation}\label{eq-nj1}
\begin{split} 
 n_e=\frac{8\pi\sqrt2}{(2\pi\hbar)^3}m_e^3c^3\beta_e^{3/2}\left[F_{1/2}(\xi,\beta_e)+\beta_e F_{3/2}(\xi,\beta_e) \right],
\end{split}
\end{equation}
where $F_k$ is the relativistic $k$-th order Fermi-Dirac integral, given by,
\begin{equation}\label{eq-FD-int}
 F_{k}(\xi,\beta_e)=\int_{0}^{\infty}\frac{\zeta^k\sqrt{1+(\beta_e/2)\zeta}}{1+\exp(\zeta-\xi)}d\zeta,
 \end{equation}
in which $ \beta_e=k_{\mathrm B}T_e/m_ec^2$ is the relativity parameter defined before, $ \zeta= E(p_e)/k_{\mathrm B}T_e$, and the normalized chemical potential energy $\xi={\mu}/k_{\mathrm B}T_e$ is called the electron degeneracy parameter. 
\par 
The electron degeneracy pressure at finite temperature $(T\neq 0~K)$  is given by \cite{boshkayev2016equilibrium,timmes1999accuracy}
\begin{equation}\label{eq-pj0}
P_e=\frac{1}{3\pi^2 \hbar^3}\int_{0}^{\infty} \frac{p_e^3}{\exp  \left(\frac{ E(p_e)- \mu}{k_{\mathrm B}T_e}\right)+1} dE,
\end{equation}
which can be expressed as
\begin{equation}\label{eq-pj1}
\begin{split}
{P}_e=\frac{2^{3/2}}{3\pi^2\hbar^3}m_e^4c^5\beta_e^{5/2}\left[F_{3/2}(\xi,\beta_e)+\frac{\beta_e}{2} F_{5/2}(\xi,\beta_e) \right].
\end{split}
\end{equation}
 Furthermore, $\xi$ satisfies the following strict condition (without the relativistic and electrostatic potential energies) 
\begin{equation}
\sum_{j=e,p}\left[1+\exp\left(-\xi\right) \right]^{-1}\leq1.
\end{equation}
Next, evaluating the integrals $F_k(\xi,\beta_e)$ in Eq. \eqref{eq-nj1} and following   the method by Landau and Lifshitz \cite{landau2013statistical}, we obtain the following  expressions for the dimensionless electron number density $n_e$ and the  partially degenerate pressure $P_e$  in two different cases of $\beta_e<1$ and $\beta_e>1$ as (See, for details, Ref. \cite{dey2024ion})
\begin{widetext}
\begin{equation} \label{eq-ne2}
n_e=
\begin{cases}
\begin{split}
 &A_e\left[ \left\lbrace\xi^{3/2}+\frac{\pi^2}{8}\xi^{-1/2}  +\frac{7\pi^4}{640}\xi^{-5/2}\right\rbrace+\frac{3\beta_e}{5}\left\lbrace\xi^{5/2}
 +\frac{5\pi^2}{8}\xi^{1/2} -\frac{7\pi^4}{384}\xi^{-3/2}\right\rbrace \right], & \text{for $\beta_e<1$},\\ \\
 &A_e\left[\Bigl\{\xi^2+\frac{\pi^2}{3} \Bigl\} +\frac{2\beta_e}{3}\Bigl\{\xi^3 +\pi^2 \xi\Bigl\} \right],& \text{for $\beta_e>1$},
\end{split}
\end{cases}
\end{equation}
\begin{equation} \label{eq-Pe2}
P_{e}=
\begin{cases}
\begin{split}
 &\frac{2}{5}\beta_eA_e\left[ \left\lbrace\xi^{5/2}+\frac{5\pi^2}{8}\xi^{1/2}  -\frac{7\pi^4}{384}\xi^{-3/2}\right\rbrace+\frac{5 \beta_e}{14}\left\lbrace\xi^{7/2}
 +\frac{35\pi^2}{24}\xi^{3/2} +\frac{49\pi^4}{384}\xi^{-1/2}\right\rbrace \right], & \text{for $\beta_e<1$},\\ \\
 &\frac{4}{9}\beta_eA_e\left[\Bigl\{\xi^3+\pi^2\xi \Bigl\} +\frac{3\beta_e}{8}\Bigl\{\xi^4 +2\pi^2 \xi^2 +\frac{7\pi^4}{15}\Bigl\} \right],& \text{for $\beta_e>1$},
\end{split}
\end{cases}
\end{equation}
where the coefficient $A_e$ is given by
\begin{equation} \label{expression-Ae}
A_e=
\begin{cases}
\begin{split}
&\left[\left(\xi_0^{3/2}+\frac{\pi^2}{8}\xi_0^{-1/2}+\frac{7\pi^4}{640}\xi_0^{-5/2}\right)+\frac{3\beta_e}{5}\left(\xi_0^{5/2}+\frac{5\pi^2}{8}\xi_0^{1/2}-\frac{7\pi^4}{384}\xi_0^{-3/2}\right)\right]^{-1},  & \text{for $\beta_j<1$}, \\ \\
 &\left[\left(\xi_0^{2}+\frac{\pi^2}{3}\right) +\frac{2\beta_e}{3}\left(\xi_0^{3}+\pi^2\xi_0\right)\right]^{-1},& \text{for $\beta_j>1$}.
\end{split}
\end{cases}
\end{equation}
\end{widetext}
The total energy density and the enthalpy for electrons are defined as ${\cal E}_e\approx n_e\sqrt{v_e^2+1}$ and $H_e\equiv n_e\sqrt{v_e^2+1}+P_e$ so that at equilibrium, $H_{e0}(x)=n_{e0}(x)(\sqrt{v_{e0}^2(x)+1}+ P_{e0}(0))$, $\alpha_e=H_{e0}(x)/{n_{e0}(x)}\sim (1+P_{e0}(0))$. The expressions for $P_{e0}(0)$ and $b$ are  
\begin{widetext}
\begin{equation}
P_{e0}(0)=
\begin{cases}
\begin{split}
 & \frac{2}{5}\beta_e A_e\left[\left\lbrace \xi_0^{5/2}+\frac{5\pi^2}{8}\xi_0^{1/2}-\frac{7\pi^4}{384}\xi_0^{-3/2}  \right\rbrace+\frac{5\beta_e }{14}\left\lbrace \xi_0^{7/2}
 +\frac{35\pi^2}{24}\xi_0^{3/2}+\frac{49\pi^4}{384}\xi_0^{-1/2} \right\rbrace\right], & \text{for $\beta_e<1$},\\ \\
 &\frac{4}{9}\beta_e A_e\left[\left\lbrace \xi_0^3+\pi^2\xi_0  \right\rbrace+\frac{3\beta_e}{8}\left\lbrace \xi_0^4+
 2\pi^2 \xi_0^2+\frac{7\pi^4}{15} \right\rbrace\right],& \text{for $\beta_e>1$},
\end{split}
\end{cases}\label{exp-pre-equ-AD)}
\end{equation}
\begin{equation} 
b=P_{e1}/n_{e1}\approx
\begin{cases}
\begin{split}
 & \frac{2\beta_e}{3}\frac{\left\lbrace \xi_0^{3/2}+\frac{\pi^2}{8}\xi_0^{-1/2}  \right\rbrace+\frac{\beta_e }{2}\left\lbrace \xi_0^{5/2}
 +\frac{5\pi^2}{8}\xi_0^{1/2} \right\rbrace}{\left\lbrace \xi_0^{1/2}-\frac{\pi^2}{24}\xi_0^{-3/2}  \right\rbrace+ {\beta_e }\left\lbrace \xi_0^{3/2}
 +\frac{\pi^2}{8}\xi_0^{-1/2} \right\rbrace} , & \text{for $\beta_e<1$},\\ \\
 &\frac{2\beta_e}{3}\frac{\left\lbrace \xi_0^2+\frac{\pi^2}{3}  \right\rbrace+\frac{\beta_e}{2}\left\lbrace \xi_0^3+
 \pi^2 \xi_0 \right\rbrace}{\left[ \xi_0+ {\beta_e }\lbrace\xi_0^2+ \frac{\pi^2}{3}\rbrace\right]},& \text{for $\beta_e>1$}.
\end{split}
\end{cases}\label{exp-b-AD}
\end{equation}
\end{widetext} 

Next, we study the RTI in two different cases, namely $\beta_e<1$ and $\beta_e>1$ as follows: 
\subsubsection*{Case I: $\beta_e<1$.} \label{sec-PD-case1}

\begin{figure} 
\centering
\includegraphics[width=3.8in,height=2.5in]{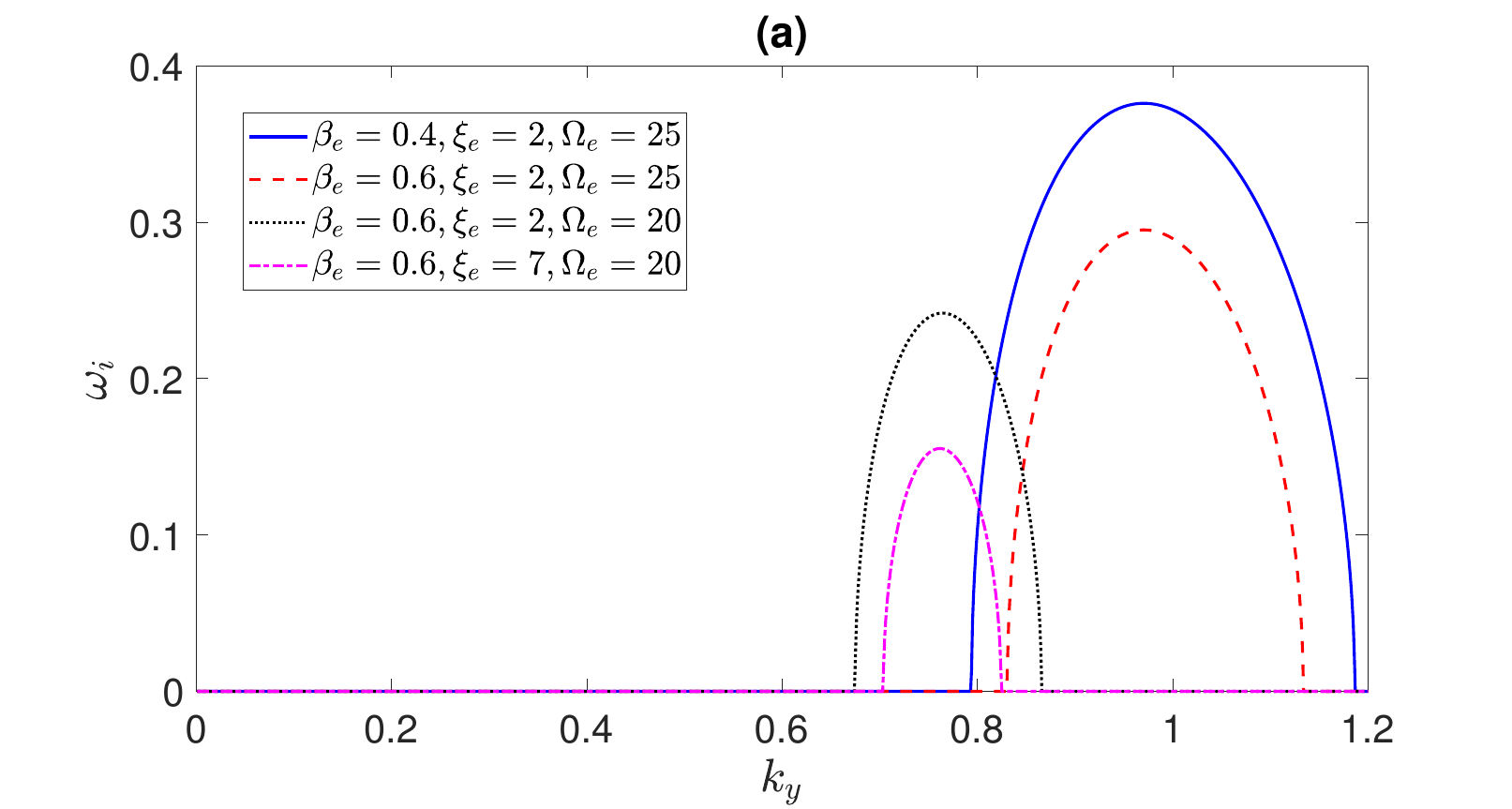}
\includegraphics[width=3.8in,height=2.5in]{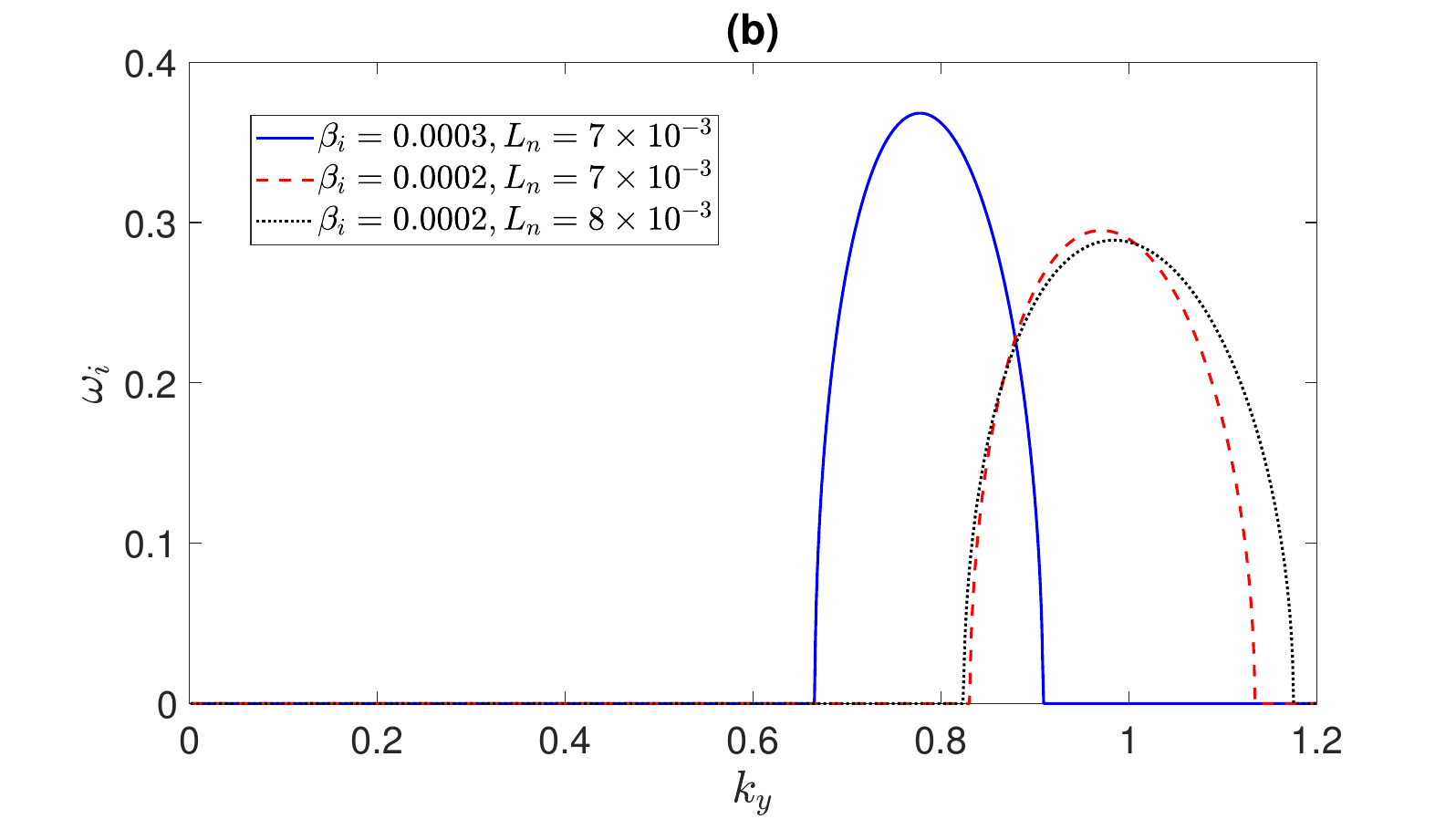}
\caption{Partially degenerate plasmas with $\beta_e<1$. The profiles of the growth rates are shown with the variations of different plasma parameters as in the legends with $m=1900$, $\Gamma=1.5$ and $\tilde{g}=2.80\times 10^{-21}$. The other fixed parameter values for subplots (a) and (b), respectively, are  ($L_n=7\times10^{-3}$ and $\beta_i=0.0002$) and ($\beta_e=0.6, ~ \xi=2 $ and $\Omega_e=25$).}
\label{fig:fig_ar_be_le_1}
\end{figure}

In this case, we consider the typical parametric values relevant for astrophysical plasmas   \cite{peterson2021effects,saumon20221current} as $n_0(0)\simeq 10^{29}-10^{31}~{\rm  cm^{-3}},~T_i\simeq 1.08 \times 10^5-4.35\times 10^9~ {\rm K},~T_e\simeq 1.78 \times 10^9-4.74\times 10^9~ {\rm K},~B_0\simeq 2.36\times 10^{9}-2.24\times 10^{12} ~{\rm G},~g\simeq 10^8 ~{\rm cm~ s^{-2}},~L_n \simeq 2.27\times 10^{-11}-4.5\times10^{-10} ~{\rm cm} $, and $\mu\simeq 1.63\times10^{-7}-1.31\times 10^{-6}~{\rm erg}$.    
 The qualitative features of the growth rate of instability by the effects of $\beta_e$, $\beta_i$, $\xi_e$, $\Omega_e$, and $L_n$ remain similar to the classical case (\textit{cf}. Fig. \ref{fig:cl_be_le_1_1}). However, the reduction of the growth rate and the expansion of the instability domain are noticeable with a small increment of the parameter $\beta_e$ and a decrement of $\beta_i$. As before, the magnetic field strength increases both the growth rate and the instability domain. In comparison with the cases of classical (Sec. \ref{sec-classical}) and fully degenerate (Sec. \ref{sec-FDC}) plasmas, an  interesting and distinct feature is noted by the effects of $\xi_e$. As the latter is increased or the electron chemical potential energy is enhanced compared to the electron rest-mass energy, the growth rate and the instability domains are significantly reduced [See subplot (a) of Fig.   \ref{fig:fig_ar_be_le_1}]. Physically, since an enhancement of the chemical energy corresponds to a decrement of the electron thermal energy, the drift velocities are reduced and so the perturbed ${\bm E}\times {\bm B}_0$-drift velocity can have upward direction to move the plama layer upward. This leads to a reduction of spikes in the plasma penetrating the lighter fluid (magnetic field), which supports the plasma, and consequently, the growth rate is reduced. Subplot (b) shows that similar to the cases of classical and  fully degenerate electrons, not only are the growth rates significantly reduced, the instability domains shift  towards higher values of $k_y$ as $\beta_i$ is slightly reduced or the length scale of inhomogeneity $L_n$ is increased. In particular, for $\beta_i\sim 0.0002,~\beta_e\sim0.4, ~\xi_e\sim 2,~ \Omega_e\sim 25$, and $L_n\sim7\times10^{-3}$, the maximum growth rate can be observed as $\omega^{\max}_{i}=0.38 $ at $k_y=0.97$. 
\subsubsection*{Case II: $\beta_e>1$.} \label{sec-PD-case2}
\begin{figure} 
\centering
\includegraphics[width=3.8in,height=2.5in]{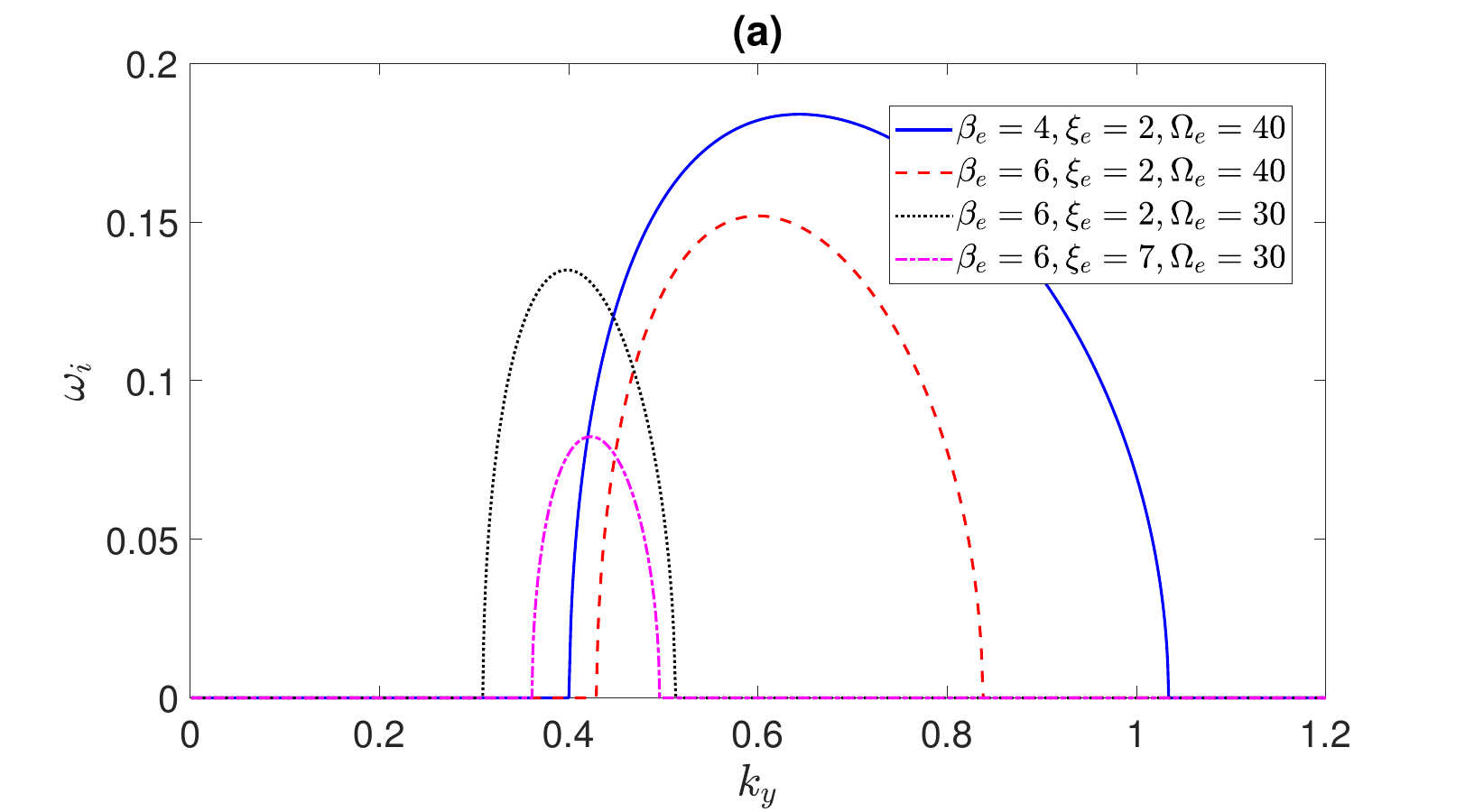}
\includegraphics[width=3.8in,height=2.5in]{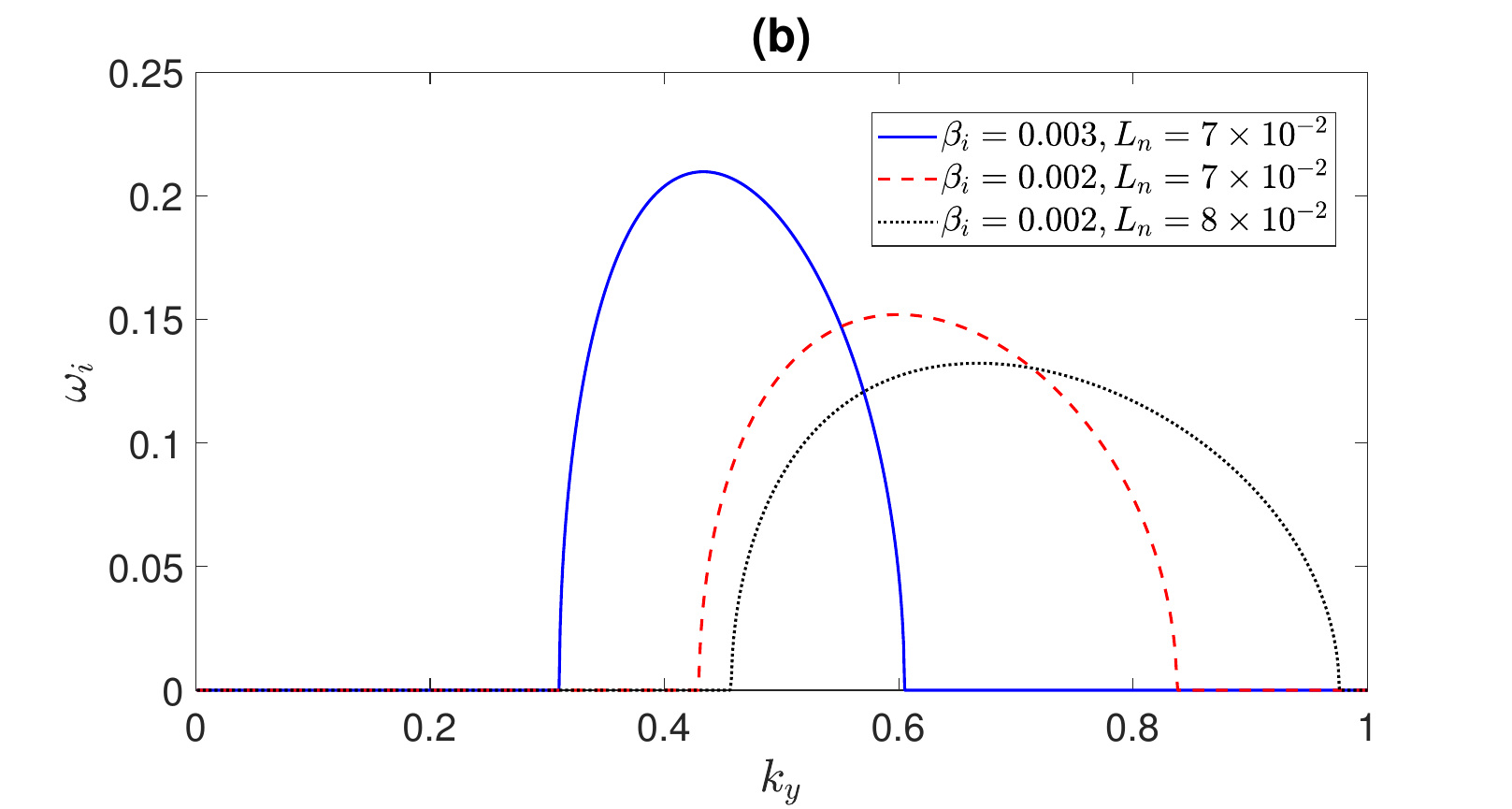}
\caption{Partially degenerate plasmas $\beta_e>1$. The profiles of the growth rates are shown with the variations of different plasma parameters as in the legends with $m=1900$, $\Gamma=1.5$ and $\tilde{g}=2.80\times 10^{-21}$. The other fixed parameter values for subplots (a) and (b), respectively, are  ($L_n=7\times10^{-2}$ and $\beta_i=0.002$) and ($\beta_e=6, ~ \xi=2 $ and $\Omega_e=40$). }
\label{fig:fig_ar_be_gt_1}
\end{figure}

In the case of higher thermal energy than the rest mass energy of electrons, we choose the typical parametric values  relevant for astrophysical plasmas \cite{peterson2021effects,saumon20221current} as  $n_0(0)\simeq 10^{30}-10^{32}~{\rm cm^{-3}},~T_i\simeq 1.08 \times 10^6-4.35\times 10^{10}~ {\rm K},~T_e\simeq 1.78 \times 10^{10}-4.74\times 10^{10}~ {\rm K},~B_0\simeq 7.38\times 10^{9}-4.11\times 10^{12} ~{\rm G},~g\simeq 10^8 ~{\rm cm~ s^{-2}},~L_n \simeq 1.13\times 10^{-10}-1.82\times10^{-8} ~{\rm cm} $, and $\mu\simeq 1.63\times10^{-6}-1.31\times 10^{-5}~{\rm erg}$. 
Here, a reduction of the growth rate is observed similar to Case I, however, in contrast to Case I, the instability domains get reduced with increasing values of $\beta_e>1$ and $\xi_e$. Also, similar to Case I, the growth rate is enhanced but the instability domain shrinks at increasing values of $\beta_i$, and both are increased at increasing values of the magnetic field strength [See subplot (a) of Fig. \ref{fig:fig_ar_be_gt_1}].  Also, the growth rate of instability gets significantly reduced but the instability domain expands and shifts towards higher values of $k_y$ by the effects of the inhomogeneity length scale $L_n$. 
[See subplot (b) of Fig. \ref{fig:fig_ar_be_gt_1}]. Thus, it may be concluded that higher electron chemical energy reduces the instability domain and the growth rate, i.e., the finite temperature degeneracy of electrons also tends to weaken the RTI. In particular,  for $\beta_i\sim 0.002, ~\xi_e\sim 2,~ \Omega_e\sim 40$, and $ L_n\sim7\times10^{-2}$, the maximum growth rate is obtained as $\omega^{\max}_{i}=0.18$ at $k_y=0.64$.  
\section{Conclusion}\label{sec-conclusion}
We have investigated the Rayleigh-Taylor instability of electrostatic perturbations in relativistic electron-ion magnetoplasmas under the influences of the unperturbed plasma density gradient and the constant gravity force. Specifically, we have focused on three different plasma regimes, namely, (i) when electrons are in isothermal equilibrium and follow the classical isothermal equation of state, (ii) electrons form a fully degenerate Fermi gas at zero temperature, and (iii) electrons are partially degenerate or have finite temperature degeneracy. Starting from a relativistic fluid model for electrons and ions and considering appropriate equations of state for classical and degenerate electrons, we have obtained the general dispersion relation, which we have analyzed in the three cases. In regimes of classical plasmas such as those in laboratory and magnetic confinement fusion plasmas, while the isothermal motion of electrons tends to reduce the instability growth rate, the ion thermal energy, and the external magnetic field enhance the RTI growth rate. Also, in regimes of higher electron thermal energy than its rest mass energy, the instability growth rate can be reduced with higher values of the length scale of density inhomogeneity than the ion plasma skin depth.   In regimes of fully degenerate plasmas (or partially degenerate plasmas), such as those in the environments of massive white dwarf stars, the effect of the electron degeneracy pressure (or the electron chemical energy) is to reduce both the instability domain and the growth rate of RTI, implying that relativistic degeneracy or finite temperature degeneracy of electrons tends to suppress the RTI and a complete suppression may occur in the ultra-relativistic regime or in the regime of higher chemical energies than the rest mass energy. The prevention of such instabilities can lead to several important insights for understanding the underlying dynamics of RTIs.
\par 
In certain astrophysical plasmas, when the electron temperature drops below $10^9$ K but higher than $10^7$ K, electrons deviate from thermal and chemical equilibrium to form a partially degenerate gas, and, since they can strongly scatter with the plasma, their distributions can be governed by the Fermi-Dirac pressure law with the energy spectrum ranging from the thermal to Fermi energies. In these partially degenerate regimes, we have seen that higher chemical energy (than the thermal energy) and higher thermal energy (than the rest-mass energy) of electrons are required to bring  both the instability domain and the growth rate of RTI.    
\par 
To conclude, the present theory should help understand the occurrence of RTI and the instability domains in a wide range of plasma environments ranging from laboratory to astrophysical settings where electrons can be in isothermal equilibrium and can deviate from the thermal and chemical equilibrium to form a partially degenerate or fully degenerate Fermi gas.   In the present work, we have neglected the effects of fluid kinematic viscosity and particle collision. Such dissipative effects will certainly modify the instability criteria and can weaken the RTI growth.  However, a detailed analysis is beyond the scope of the present work.  
\section*{ACKNOWLEDGMENTS}
One of us, RD, acknowledges support from the University Grants Commission (UGC), Government of India, for a Senior
Research Fellowship (SRF) with Ref. No. 1161/(CSIR-UGC NET DEC. 2018) and F. no. 16-6 (DEC. 2018)/2019 (NET/CSIR).
\section*{AUTHOR DECLARATIONS}
\subsection*{Conflict of Interest}
The authors have no conﬂicts to disclose.
\subsection*{Author Contributions}
Rupak Dey: Formal analysis (equal); Investigation (equal); Methodology (equal); Writing--original draft (equal). Amar Prasad
Misra: Conceptualization (equal); Investigation (equal); Methodology
(equal); Supervision (equal); Validation (equal); Writing--review \& editing (equal).
\section*{DATA AVAILABILITY}
The data that support the ﬁndings of this study are available from the corresponding author upon reasonable request.
\bibliography{ref}

\begin{thebibliography}{45}%
\makeatletter
\providecommand \@ifxundefined [1]{%
 \@ifx{#1\undefined}
}%
\providecommand \@ifnum [1]{%
 \ifnum #1\expandafter \@firstoftwo
 \else \expandafter \@secondoftwo
 \fi
}%
\providecommand \@ifx [1]{%
 \ifx #1\expandafter \@firstoftwo
 \else \expandafter \@secondoftwo
 \fi
}%
\providecommand \natexlab [1]{#1}%
\providecommand \enquote  [1]{``#1''}%
\providecommand \bibnamefont  [1]{#1}%
\providecommand \bibfnamefont [1]{#1}%
\providecommand \citenamefont [1]{#1}%
\providecommand \href@noop [0]{\@secondoftwo}%
\providecommand \href [0]{\begingroup \@sanitize@url \@href}%
\providecommand \@href[1]{\@@startlink{#1}\@@href}%
\providecommand \@@href[1]{\endgroup#1\@@endlink}%
\providecommand \@sanitize@url [0]{\catcode `\\12\catcode `\$12\catcode
  `\&12\catcode `\#12\catcode `\^12\catcode `\_12\catcode `\%12\relax}%
\providecommand \@@startlink[1]{}%
\providecommand \@@endlink[0]{}%
\providecommand \url  [0]{\begingroup\@sanitize@url \@url }%
\providecommand \@url [1]{\endgroup\@href {#1}{\urlprefix }}%
\providecommand \urlprefix  [0]{URL }%
\providecommand \Eprint [0]{\href }%
\providecommand \doibase [0]{https://doi.org/}%
\providecommand \selectlanguage [0]{\@gobble}%
\providecommand \bibinfo  [0]{\@secondoftwo}%
\providecommand \bibfield  [0]{\@secondoftwo}%
\providecommand \translation [1]{[#1]}%
\providecommand \BibitemOpen [0]{}%
\providecommand \bibitemStop [0]{}%
\providecommand \bibitemNoStop [0]{.\EOS\space}%
\providecommand \EOS [0]{\spacefactor3000\relax}%
\providecommand \BibitemShut  [1]{\csname bibitem#1\endcsname}%
\let\auto@bib@innerbib\@empty
\bibitem [{\citenamefont {Maryam}\ \emph {et~al.}(2021)\citenamefont {Maryam},
  \citenamefont {Rozina},\ and\ \citenamefont {Ali}}]{maryam2021rayleigh}%
  \BibitemOpen
  \bibfield  {author} {\bibinfo {author} {\bibfnamefont {N.}~\bibnamefont
  {Maryam}}, \bibinfo {author} {\bibfnamefont {C.}~\bibnamefont {Rozina}},\
  and\ \bibinfo {author} {\bibfnamefont {S.}~\bibnamefont {Ali}},\ }\bibfield
  {title} {\bibinfo {title} {Rayleigh–taylor instability in electron–ion
  radiative dense plasmas},\ }\href {https://doi.org/10.1109/TPS.2021.3055478}
  {\bibfield  {journal} {\bibinfo  {journal} {IEEE Transactions on Plasma
  Science}\ }\textbf {\bibinfo {volume} {49}},\ \bibinfo {pages} {1072}
  (\bibinfo {year} {2021})}\BibitemShut {NoStop}%
\bibitem [{\citenamefont {Bhambhu}\ and\ \citenamefont
  {Prajapati}(2024)}]{bhambhu2024radiation}%
  \BibitemOpen
  \bibfield  {author} {\bibinfo {author} {\bibfnamefont {R.}~\bibnamefont
  {Bhambhu}}\ and\ \bibinfo {author} {\bibfnamefont {R.~P.}\ \bibnamefont
  {Prajapati}},\ }\bibfield  {title} {\bibinfo {title} {{Radiation
  pressure-driven Rayleigh–Taylor instability in compressible strongly
  magnetized ultra-relativistic degenerate plasmas}},\ }\href
  {https://doi.org/10.1063/5.0209141} {\bibfield  {journal} {\bibinfo
  {journal} {Physics of Plasmas}\ }\textbf {\bibinfo {volume} {31}},\ \bibinfo
  {pages} {082708} (\bibinfo {year} {2024})}\BibitemShut {NoStop}%
\bibitem [{\citenamefont {Garai}\ \emph {et~al.}(2020)\citenamefont {Garai},
  \citenamefont {Ghose-Choudhury},\ and\ \citenamefont
  {Guha}}]{garai2020rayleigh}%
  \BibitemOpen
  \bibfield  {author} {\bibinfo {author} {\bibfnamefont {S.}~\bibnamefont
  {Garai}}, \bibinfo {author} {\bibfnamefont {A.}~\bibnamefont
  {Ghose-Choudhury}},\ and\ \bibinfo {author} {\bibfnamefont {P.}~\bibnamefont
  {Guha}},\ }\bibfield  {title} {\bibinfo {title} {Rayleigh taylor like
  instability in presence of shear velocity in a strongly coupled quantum
  plasma},\ }\href {https://doi.org/10.1088/1402-4896/abb697} {\bibfield
  {journal} {\bibinfo  {journal} {Physica Scripta}\ }\textbf {\bibinfo {volume}
  {95}},\ \bibinfo {pages} {105605} (\bibinfo {year} {2020})}\BibitemShut
  {NoStop}%
\bibitem [{\citenamefont
  {Chandrasekhar}(1981)}]{chandrasekhar1961hydrodynamic}%
  \BibitemOpen
  \bibfield  {author} {\bibinfo {author} {\bibfnamefont {S.}~\bibnamefont
  {Chandrasekhar}},\ }\href@noop {} {\emph {\bibinfo {title} {Hydrodynamic and
  Hydromagnetic Stability}}},\ Dover Books on Physics Series\ (\bibinfo
  {publisher} {Dover Publications},\ \bibinfo {year} {1981})\BibitemShut
  {NoStop}%
\bibitem [{\citenamefont {Onishchenko}\ \emph {et~al.}(2011)\citenamefont
  {Onishchenko}, \citenamefont {Pokhotelov}, \citenamefont {Stenflo},\ and\
  \citenamefont {Shukla}}]{onishchenko2011magnetic}%
  \BibitemOpen
  \bibfield  {author} {\bibinfo {author} {\bibfnamefont {O.}~\bibnamefont
  {Onishchenko}}, \bibinfo {author} {\bibfnamefont {O.}~\bibnamefont
  {Pokhotelov}}, \bibinfo {author} {\bibfnamefont {L.}~\bibnamefont
  {Stenflo}},\ and\ \bibinfo {author} {\bibfnamefont {P.}~\bibnamefont
  {Shukla}},\ }\bibfield  {title} {\bibinfo {title} {The magnetic
  rayleigh--taylor instability and flute waves at the ion larmor radius
  scales},\ }\href@noop {} {\bibfield  {journal} {\bibinfo  {journal} {Physics
  of Plasmas}\ }\textbf {\bibinfo {volume} {18}} (\bibinfo {year}
  {2011})}\BibitemShut {NoStop}%
\bibitem [{\citenamefont {Betti}\ and\ \citenamefont
  {Hurricane}(2016)}]{betti2016inertial}%
  \BibitemOpen
  \bibfield  {author} {\bibinfo {author} {\bibfnamefont {R.}~\bibnamefont
  {Betti}}\ and\ \bibinfo {author} {\bibfnamefont {O.}~\bibnamefont
  {Hurricane}},\ }\bibfield  {title} {\bibinfo {title} {Inertial-confinement
  fusion with lasers},\ }\href@noop {} {\bibfield  {journal} {\bibinfo
  {journal} {Nature Physics}\ }\textbf {\bibinfo {volume} {12}},\ \bibinfo
  {pages} {435} (\bibinfo {year} {2016})}\BibitemShut {NoStop}%
\bibitem [{\citenamefont {Tanaka}\ \emph {et~al.}(2018)\citenamefont {Tanaka},
  \citenamefont {Toma},\ and\ \citenamefont
  {Tominaga}}]{tanaka2018confinement}%
  \BibitemOpen
  \bibfield  {author} {\bibinfo {author} {\bibfnamefont {S.~J.}\ \bibnamefont
  {Tanaka}}, \bibinfo {author} {\bibfnamefont {K.}~\bibnamefont {Toma}},\ and\
  \bibinfo {author} {\bibfnamefont {N.}~\bibnamefont {Tominaga}},\ }\bibfield
  {title} {\bibinfo {title} {Confinement of the crab nebula with tangled
  magnetic field by its supernova remnant},\ }\href@noop {} {\bibfield
  {journal} {\bibinfo  {journal} {Monthly Notices of the Royal Astronomical
  Society}\ }\textbf {\bibinfo {volume} {478}},\ \bibinfo {pages} {4622}
  (\bibinfo {year} {2018})}\BibitemShut {NoStop}%
\bibitem [{\citenamefont {Duffell}\ and\ \citenamefont
  {Kasen}(2017)}]{duffell2017rayleigh}%
  \BibitemOpen
  \bibfield  {author} {\bibinfo {author} {\bibfnamefont {P.~C.}\ \bibnamefont
  {Duffell}}\ and\ \bibinfo {author} {\bibfnamefont {D.}~\bibnamefont
  {Kasen}},\ }\bibfield  {title} {\bibinfo {title} {Rayleigh--taylor
  instability in interacting supernovae: Implications for synchrotron magnetic
  fields},\ }\href@noop {} {\bibfield  {journal} {\bibinfo  {journal} {The
  Astrophysical Journal}\ }\textbf {\bibinfo {volume} {842}},\ \bibinfo {pages}
  {18} (\bibinfo {year} {2017})}\BibitemShut {NoStop}%
\bibitem [{\citenamefont {Rigon}\ \emph {et~al.}(2019)\citenamefont {Rigon},
  \citenamefont {Casner}, \citenamefont {Albertazzi}, \citenamefont {Michel},
  \citenamefont {Mabey}, \citenamefont {Falize}, \citenamefont {Ballet},
  \citenamefont {Som}, \citenamefont {Pikuz}, \citenamefont {Sakawa} \emph
  {et~al.}}]{rigon2019rayleigh}%
  \BibitemOpen
  \bibfield  {author} {\bibinfo {author} {\bibfnamefont {G.}~\bibnamefont
  {Rigon}}, \bibinfo {author} {\bibfnamefont {A.}~\bibnamefont {Casner}},
  \bibinfo {author} {\bibfnamefont {B.}~\bibnamefont {Albertazzi}}, \bibinfo
  {author} {\bibfnamefont {T.}~\bibnamefont {Michel}}, \bibinfo {author}
  {\bibfnamefont {P.}~\bibnamefont {Mabey}}, \bibinfo {author} {\bibfnamefont
  {E.}~\bibnamefont {Falize}}, \bibinfo {author} {\bibfnamefont
  {J.}~\bibnamefont {Ballet}}, \bibinfo {author} {\bibfnamefont {L.~V.~B.}\
  \bibnamefont {Som}}, \bibinfo {author} {\bibfnamefont {S.}~\bibnamefont
  {Pikuz}}, \bibinfo {author} {\bibfnamefont {Y.}~\bibnamefont {Sakawa}}, \emph
  {et~al.},\ }\bibfield  {title} {\bibinfo {title} {Rayleigh-taylor instability
  experiments on the luli2000 laser in scaled conditions for young supernova
  remnants},\ }\href@noop {} {\bibfield  {journal} {\bibinfo  {journal}
  {Physical Review E}\ }\textbf {\bibinfo {volume} {100}},\ \bibinfo {pages}
  {021201} (\bibinfo {year} {2019})}\BibitemShut {NoStop}%
\bibitem [{\citenamefont {Peterson}\ \emph {et~al.}(2021)\citenamefont
  {Peterson}, \citenamefont {Dexheimer}, \citenamefont {Negreiros},\ and\
  \citenamefont {Castanheira}}]{peterson2021effects}%
  \BibitemOpen
  \bibfield  {author} {\bibinfo {author} {\bibfnamefont {J.}~\bibnamefont
  {Peterson}}, \bibinfo {author} {\bibfnamefont {V.}~\bibnamefont {Dexheimer}},
  \bibinfo {author} {\bibfnamefont {R.}~\bibnamefont {Negreiros}},\ and\
  \bibinfo {author} {\bibfnamefont {B.~G.}\ \bibnamefont {Castanheira}},\
  }\bibfield  {title} {\bibinfo {title} {Effects of magnetic fields in hot
  white dwarfs},\ }\href {https://doi.org/10.3847/1538-4357/ac1ba7} {\bibfield
  {journal} {\bibinfo  {journal} {The Astrophysical Journal}\ }\textbf
  {\bibinfo {volume} {921}},\ \bibinfo {pages} {1} (\bibinfo {year}
  {2021})}\BibitemShut {NoStop}%
\bibitem [{\citenamefont {Saumon}\ \emph {et~al.}(2022)\citenamefont {Saumon},
  \citenamefont {Blouin},\ and\ \citenamefont {Tremblay}}]{saumon20221current}%
  \BibitemOpen
  \bibfield  {author} {\bibinfo {author} {\bibfnamefont {D.}~\bibnamefont
  {Saumon}}, \bibinfo {author} {\bibfnamefont {S.}~\bibnamefont {Blouin}},\
  and\ \bibinfo {author} {\bibfnamefont {P.-E.}\ \bibnamefont {Tremblay}},\
  }\bibfield  {title} {\bibinfo {title} {Current challenges in the physics of
  white dwarf stars},\ }\href
  {https://doi.org/https://doi.org/10.1016/j.physrep.2022.09.001} {\bibfield
  {journal} {\bibinfo  {journal} {Physics Reports}\ }\textbf {\bibinfo {volume}
  {988}},\ \bibinfo {pages} {1} (\bibinfo {year} {2022})},\ \bibinfo {note}
  {current Challenges in the Physics of White Dwarf Stars}\BibitemShut
  {NoStop}%
\bibitem [{\citenamefont {{Bejger, M.}}\ and\ \citenamefont {{Haensel,
  P.}}(2004)}]{bejger2004surface}%
  \BibitemOpen
  \bibfield  {author} {\bibinfo {author} {\bibnamefont {{Bejger, M.}}}\ and\
  \bibinfo {author} {\bibnamefont {{Haensel, P.}}},\ }\bibfield  {title}
  {\bibinfo {title} {Surface gravity of neutron stars and strange stars},\
  }\href {https://doi.org/10.1051/0004-6361:20034538} {\bibfield  {journal}
  {\bibinfo  {journal} {Astronomy \& Astrophysics}\ }\textbf {\bibinfo {volume}
  {420}},\ \bibinfo {pages} {987} (\bibinfo {year} {2004})}\BibitemShut
  {NoStop}%
\bibitem [{\citenamefont {Lasky}(2015)}]{lasky2015gravitational}%
  \BibitemOpen
  \bibfield  {author} {\bibinfo {author} {\bibfnamefont {P.~D.}\ \bibnamefont
  {Lasky}},\ }\bibfield  {title} {\bibinfo {title} {Gravitational waves from
  neutron stars: A review},\ }\href {https://doi.org/10.1017/pasa.2015.35}
  {\bibfield  {journal} {\bibinfo  {journal} {Publications of the Astronomical
  Society of Australia}\ }\textbf {\bibinfo {volume} {32}},\ \bibinfo {pages}
  {e034} (\bibinfo {year} {2015})}\BibitemShut {NoStop}%
\bibitem [{\citenamefont {Zhu}\ \emph {et~al.}(2023)\citenamefont {Zhu},
  \citenamefont {Stone},\ and\ \citenamefont {Calvet}}]{zhu2023global}%
  \BibitemOpen
  \bibfield  {author} {\bibinfo {author} {\bibfnamefont {Z.}~\bibnamefont
  {Zhu}}, \bibinfo {author} {\bibfnamefont {J.~M.}\ \bibnamefont {Stone}},\
  and\ \bibinfo {author} {\bibfnamefont {N.}~\bibnamefont {Calvet}},\
  }\bibfield  {title} {\bibinfo {title} {A global 3-d simulation of
  magnetospheric accretion: I. magnetically disrupted discs and surface
  accretion},\ }\href@noop {} {\bibfield  {journal} {\bibinfo  {journal}
  {Monthly Notices of the Royal Astronomical Society}\ ,\ \bibinfo {pages}
  {stad3712}} (\bibinfo {year} {2023})}\BibitemShut {NoStop}%
\bibitem [{\citenamefont {Yang}\ \emph {et~al.}(2019)\citenamefont {Yang},
  \citenamefont {Rallabandi},\ and\ \citenamefont
  {Stone}}]{yang2019autophoresis}%
  \BibitemOpen
  \bibfield  {author} {\bibinfo {author} {\bibfnamefont {F.}~\bibnamefont
  {Yang}}, \bibinfo {author} {\bibfnamefont {B.}~\bibnamefont {Rallabandi}},\
  and\ \bibinfo {author} {\bibfnamefont {H.~A.}\ \bibnamefont {Stone}},\
  }\bibfield  {title} {\bibinfo {title} {Autophoresis of two
  adsorbing/desorbing particles in an electrolyte solution},\ }\href@noop {}
  {\bibfield  {journal} {\bibinfo  {journal} {Journal of Fluid Mechanics}\
  }\textbf {\bibinfo {volume} {865}},\ \bibinfo {pages} {440} (\bibinfo {year}
  {2019})}\BibitemShut {NoStop}%
\bibitem [{\citenamefont {Avinash}\ and\ \citenamefont
  {Sen}(2015)}]{avinash2015rayleigh}%
  \BibitemOpen
  \bibfield  {author} {\bibinfo {author} {\bibfnamefont {K.}~\bibnamefont
  {Avinash}}\ and\ \bibinfo {author} {\bibfnamefont {A.}~\bibnamefont {Sen}},\
  }\bibfield  {title} {\bibinfo {title} {Rayleigh-taylor instability in dusty
  plasma experiment},\ }\href@noop {} {\bibfield  {journal} {\bibinfo
  {journal} {Physics of Plasmas}\ }\textbf {\bibinfo {volume} {22}} (\bibinfo
  {year} {2015})}\BibitemShut {NoStop}%
\bibitem [{\citenamefont {Dolai}\ \emph {et~al.}(2016)\citenamefont {Dolai},
  \citenamefont {Prajapati},\ and\ \citenamefont
  {Chhajlani}}]{dolai2016effect}%
  \BibitemOpen
  \bibfield  {author} {\bibinfo {author} {\bibfnamefont {B.}~\bibnamefont
  {Dolai}}, \bibinfo {author} {\bibfnamefont {R.}~\bibnamefont {Prajapati}},\
  and\ \bibinfo {author} {\bibfnamefont {R.}~\bibnamefont {Chhajlani}},\
  }\bibfield  {title} {\bibinfo {title} {Effect of different dust flow
  velocities on combined kelvin-helmholtz and rayleigh-taylor instabilities in
  magnetized incompressible dusty fluids},\ }\href@noop {} {\bibfield
  {journal} {\bibinfo  {journal} {Physics of Plasmas}\ }\textbf {\bibinfo
  {volume} {23}} (\bibinfo {year} {2016})}\BibitemShut {NoStop}%
\bibitem [{\citenamefont {Sekar}\ and\ \citenamefont
  {Kherani}(1999)}]{sekar1999effects}%
  \BibitemOpen
  \bibfield  {author} {\bibinfo {author} {\bibfnamefont {R.}~\bibnamefont
  {Sekar}}\ and\ \bibinfo {author} {\bibfnamefont {E.}~\bibnamefont
  {Kherani}},\ }\bibfield  {title} {\bibinfo {title} {Effects of molecular ions
  on the rayleigh--taylor instability in the night-time equatorial
  ionosphere},\ }\href@noop {} {\bibfield  {journal} {\bibinfo  {journal}
  {Journal of atmospheric and solar-terrestrial physics}\ }\textbf {\bibinfo
  {volume} {61}},\ \bibinfo {pages} {399} (\bibinfo {year} {1999})}\BibitemShut
  {NoStop}%
\bibitem [{\citenamefont {Yan}\ \emph {et~al.}(2023)\citenamefont {Yan},
  \citenamefont {Parks}, \citenamefont {Mozer}, \citenamefont {Goldstein},
  \citenamefont {Chen},\ and\ \citenamefont {Liu}}]{yan2023rayleigh}%
  \BibitemOpen
  \bibfield  {author} {\bibinfo {author} {\bibfnamefont {G.}~\bibnamefont
  {Yan}}, \bibinfo {author} {\bibfnamefont {G.}~\bibnamefont {Parks}}, \bibinfo
  {author} {\bibfnamefont {F.}~\bibnamefont {Mozer}}, \bibinfo {author}
  {\bibfnamefont {M.}~\bibnamefont {Goldstein}}, \bibinfo {author}
  {\bibfnamefont {T.}~\bibnamefont {Chen}},\ and\ \bibinfo {author}
  {\bibfnamefont {Y.}~\bibnamefont {Liu}},\ }\bibfield  {title} {\bibinfo
  {title} {Rayleigh-taylor instability observed at the dayside magnetopause
  under northward interplanetary magnetic field},\ }\href@noop {} {\bibfield
  {journal} {\bibinfo  {journal} {Journal of Geophysical Research: Space
  Physics}\ }\textbf {\bibinfo {volume} {128}},\ \bibinfo {pages}
  {e2023JA031461} (\bibinfo {year} {2023})}\BibitemShut {NoStop}%
\bibitem [{\citenamefont {Archer}\ \emph {et~al.}(2019)\citenamefont {Archer},
  \citenamefont {Hietala}, \citenamefont {Hartinger}, \citenamefont
  {Plaschke},\ and\ \citenamefont {Angelopoulos}}]{archer2019direct}%
  \BibitemOpen
  \bibfield  {author} {\bibinfo {author} {\bibfnamefont {M.}~\bibnamefont
  {Archer}}, \bibinfo {author} {\bibfnamefont {H.}~\bibnamefont {Hietala}},
  \bibinfo {author} {\bibfnamefont {M.~D.}\ \bibnamefont {Hartinger}}, \bibinfo
  {author} {\bibfnamefont {F.}~\bibnamefont {Plaschke}},\ and\ \bibinfo
  {author} {\bibfnamefont {V.}~\bibnamefont {Angelopoulos}},\ }\bibfield
  {title} {\bibinfo {title} {Direct observations of a surface eigenmode of the
  dayside magnetopause},\ }\href@noop {} {\bibfield  {journal} {\bibinfo
  {journal} {Nature communications}\ }\textbf {\bibinfo {volume} {10}},\
  \bibinfo {pages} {615} (\bibinfo {year} {2019})}\BibitemShut {NoStop}%
\bibitem [{\citenamefont {Adams}\ \emph {et~al.}(2015)\citenamefont {Adams},
  \citenamefont {Moser},\ and\ \citenamefont {Hsu}}]{adams2015observation}%
  \BibitemOpen
  \bibfield  {author} {\bibinfo {author} {\bibfnamefont {C.~S.}\ \bibnamefont
  {Adams}}, \bibinfo {author} {\bibfnamefont {A.~L.}\ \bibnamefont {Moser}},\
  and\ \bibinfo {author} {\bibfnamefont {S.~C.}\ \bibnamefont {Hsu}},\
  }\bibfield  {title} {\bibinfo {title} {Observation of
  rayleigh-taylor-instability evolution in a plasma with magnetic and viscous
  effects},\ }\href@noop {} {\bibfield  {journal} {\bibinfo  {journal}
  {Physical Review E}\ }\textbf {\bibinfo {volume} {92}},\ \bibinfo {pages}
  {051101} (\bibinfo {year} {2015})}\BibitemShut {NoStop}%
\bibitem [{\citenamefont {Mikhailenko}\ \emph {et~al.}(2002)\citenamefont
  {Mikhailenko}, \citenamefont {Mikhailenko},\ and\ \citenamefont
  {Weiland}}]{mikhailenko2002rayleigh}%
  \BibitemOpen
  \bibfield  {author} {\bibinfo {author} {\bibfnamefont {V.~S.}\ \bibnamefont
  {Mikhailenko}}, \bibinfo {author} {\bibfnamefont {V.~V.}\ \bibnamefont
  {Mikhailenko}},\ and\ \bibinfo {author} {\bibfnamefont {J.}~\bibnamefont
  {Weiland}},\ }\bibfield  {title} {\bibinfo {title} {Rayleigh--taylor
  instability in plasmas with shear flow},\ }\href@noop {} {\bibfield
  {journal} {\bibinfo  {journal} {Physics of Plasmas}\ }\textbf {\bibinfo
  {volume} {9}},\ \bibinfo {pages} {2891} (\bibinfo {year} {2002})}\BibitemShut
  {NoStop}%
\bibitem [{\citenamefont {Bhambhu}\ and\ \citenamefont
  {Prajapati}(2023)}]{bhambhu2023rayleigh}%
  \BibitemOpen
  \bibfield  {author} {\bibinfo {author} {\bibfnamefont {R.}~\bibnamefont
  {Bhambhu}}\ and\ \bibinfo {author} {\bibfnamefont {R.~P.}\ \bibnamefont
  {Prajapati}},\ }\bibfield  {title} {\bibinfo {title} {Rayleigh--taylor
  instability in compressible ultra-relativistic degenerate strongly coupled
  plasma},\ }\href@noop {} {\bibfield  {journal} {\bibinfo  {journal} {Physics
  of Plasmas}\ }\textbf {\bibinfo {volume} {30}} (\bibinfo {year}
  {2023})}\BibitemShut {NoStop}%
\bibitem [{\citenamefont {Misra}\ and\ \citenamefont
  {Chatterjee}(2018)}]{misra2018stimulated}%
  \BibitemOpen
  \bibfield  {author} {\bibinfo {author} {\bibfnamefont {A.}~\bibnamefont
  {Misra}}\ and\ \bibinfo {author} {\bibfnamefont {D.}~\bibnamefont
  {Chatterjee}},\ }\bibfield  {title} {\bibinfo {title} {Stimulated scattering
  instability in a relativistic plasma},\ }\href@noop {} {\bibfield  {journal}
  {\bibinfo  {journal} {Physics of Plasmas}\ }\textbf {\bibinfo {volume} {25}}
  (\bibinfo {year} {2018})}\BibitemShut {NoStop}%
\bibitem [{\citenamefont {Turi}\ and\ \citenamefont
  {Misra}(2022)}]{turi2022magnetohydrodynamic}%
  \BibitemOpen
  \bibfield  {author} {\bibinfo {author} {\bibfnamefont {J.}~\bibnamefont
  {Turi}}\ and\ \bibinfo {author} {\bibfnamefont {A.}~\bibnamefont {Misra}},\
  }\bibfield  {title} {\bibinfo {title} {Magnetohydrodynamic instabilities in a
  self-gravitating rotating cosmic plasma},\ }\href@noop {} {\bibfield
  {journal} {\bibinfo  {journal} {Physica Scripta}\ }\textbf {\bibinfo {volume}
  {97}},\ \bibinfo {pages} {125603} (\bibinfo {year} {2022})}\BibitemShut
  {NoStop}%
\bibitem [{\citenamefont {Chakrabarti}\ and\ \citenamefont
  {Kaw}(1996)}]{chakrabarti1996velocity}%
  \BibitemOpen
  \bibfield  {author} {\bibinfo {author} {\bibfnamefont {N.}~\bibnamefont
  {Chakrabarti}}\ and\ \bibinfo {author} {\bibfnamefont {P.}~\bibnamefont
  {Kaw}},\ }\bibfield  {title} {\bibinfo {title} {Velocity shear effect on
  rayleigh--taylor vortices in nonuniform magnetized plasmas},\ }\href@noop {}
  {\bibfield  {journal} {\bibinfo  {journal} {Physics of Plasmas}\ }\textbf
  {\bibinfo {volume} {3}},\ \bibinfo {pages} {3599} (\bibinfo {year}
  {1996})}\BibitemShut {NoStop}%
\bibitem [{\citenamefont {Chakrabarti}\ and\ \citenamefont
  {Spatschek}(1998)}]{chakrabarti1998rayleigh}%
  \BibitemOpen
  \bibfield  {author} {\bibinfo {author} {\bibfnamefont {N.}~\bibnamefont
  {Chakrabarti}}\ and\ \bibinfo {author} {\bibfnamefont {K.}~\bibnamefont
  {Spatschek}},\ }\bibfield  {title} {\bibinfo {title} {Rayleigh--taylor modes
  in the presence of velocity shear and vortices},\ }\href@noop {} {\bibfield
  {journal} {\bibinfo  {journal} {Journal of plasma physics}\ }\textbf
  {\bibinfo {volume} {59}},\ \bibinfo {pages} {737} (\bibinfo {year}
  {1998})}\BibitemShut {NoStop}%
\bibitem [{\citenamefont {Shahmansouri}\ and\ \citenamefont
  {Misra}(2016)}]{shahmansouri2016}%
  \BibitemOpen
  \bibfield  {author} {\bibinfo {author} {\bibfnamefont {M.}~\bibnamefont
  {Shahmansouri}}\ and\ \bibinfo {author} {\bibfnamefont {A.~P.}\ \bibnamefont
  {Misra}},\ }\bibfield  {title} {\bibinfo {title} {Modulation and nonlinear
  evolution of multi-dimensional langmuir wave envelopes in a relativistic
  plasma},\ }\href {https://doi.org/10.1063/1.4971444} {\bibfield  {journal}
  {\bibinfo  {journal} {Physics of Plasmas}\ }\textbf {\bibinfo {volume}
  {23}},\ \bibinfo {pages} {122112} (\bibinfo {year} {2016})},\ \Eprint
  {https://arxiv.org/abs/https://pubs.aip.org/aip/pop/article-pdf/doi/10.1063/1.4971444/16006750/122112\_1\_online.pdf}
  {https://pubs.aip.org/aip/pop/article-pdf/doi/10.1063/1.4971444/16006750/122112\_1\_online.pdf}
  \BibitemShut {NoStop}%
\bibitem [{\citenamefont {Shahmansouri}\ \emph {et~al.}(2017)\citenamefont
  {Shahmansouri}, \citenamefont {Alinejad},\ and\ \citenamefont
  {Tribeche}}]{shahmansouri2017}%
  \BibitemOpen
  \bibfield  {author} {\bibinfo {author} {\bibfnamefont {M.}~\bibnamefont
  {Shahmansouri}}, \bibinfo {author} {\bibfnamefont {H.}~\bibnamefont
  {Alinejad}},\ and\ \bibinfo {author} {\bibfnamefont {M.}~\bibnamefont
  {Tribeche}},\ }\bibfield  {title} {\bibinfo {title} {Breather structures in
  degenerate relativistic non-extensive plasma},\ }\href
  {https://doi.org/10.1017/S0022377817000332} {\bibfield  {journal} {\bibinfo
  {journal} {Journal of Plasma Physics}\ }\textbf {\bibinfo {volume} {83}},\
  \bibinfo {pages} {905830303} (\bibinfo {year} {2017})}\BibitemShut {NoStop}%
\bibitem [{\citenamefont {Bychkov}\ \emph {et~al.}(2008)\citenamefont
  {Bychkov}, \citenamefont {Marklund},\ and\ \citenamefont
  {Modestov}}]{bychkov2008rayleigh}%
  \BibitemOpen
  \bibfield  {author} {\bibinfo {author} {\bibfnamefont {V.}~\bibnamefont
  {Bychkov}}, \bibinfo {author} {\bibfnamefont {M.}~\bibnamefont {Marklund}},\
  and\ \bibinfo {author} {\bibfnamefont {M.}~\bibnamefont {Modestov}},\
  }\bibfield  {title} {\bibinfo {title} {The rayleigh--taylor instability and
  internal waves in quantum plasmas},\ }\href@noop {} {\bibfield  {journal}
  {\bibinfo  {journal} {Physics Letters A}\ }\textbf {\bibinfo {volume}
  {372}},\ \bibinfo {pages} {3042} (\bibinfo {year} {2008})}\BibitemShut
  {NoStop}%
\bibitem [{\citenamefont {Hoshoudy}(2009)}]{hoshoudy2009quantum}%
  \BibitemOpen
  \bibfield  {author} {\bibinfo {author} {\bibfnamefont {G.}~\bibnamefont
  {Hoshoudy}},\ }\bibfield  {title} {\bibinfo {title} {Quantum effects on the
  rayleigh--taylor instability in a horizontal inhomogeneous rotating plasma},\
  }\href@noop {} {\bibfield  {journal} {\bibinfo  {journal} {Physics of
  Plasmas}\ }\textbf {\bibinfo {volume} {16}} (\bibinfo {year}
  {2009})}\BibitemShut {NoStop}%
\bibitem [{\citenamefont {Ali}\ \emph {et~al.}(2009)\citenamefont {Ali},
  \citenamefont {Ahmed}, \citenamefont {Mirza},\ and\ \citenamefont
  {Ahmad}}]{ali2009rayleigh}%
  \BibitemOpen
  \bibfield  {author} {\bibinfo {author} {\bibfnamefont {S.}~\bibnamefont
  {Ali}}, \bibinfo {author} {\bibfnamefont {Z.}~\bibnamefont {Ahmed}}, \bibinfo
  {author} {\bibfnamefont {A.~M.}\ \bibnamefont {Mirza}},\ and\ \bibinfo
  {author} {\bibfnamefont {I.}~\bibnamefont {Ahmad}},\ }\bibfield  {title}
  {\bibinfo {title} {Rayleigh--taylor/gravitational instability in dense
  magnetoplasmas},\ }\href@noop {} {\bibfield  {journal} {\bibinfo  {journal}
  {Physics letters A}\ }\textbf {\bibinfo {volume} {373}},\ \bibinfo {pages}
  {2940} (\bibinfo {year} {2009})}\BibitemShut {NoStop}%
\bibitem [{\citenamefont {{Adak}}\ \emph {et~al.}(2014)\citenamefont {{Adak}},
  \citenamefont {{Ghosh}},\ and\ \citenamefont
  {{Chakrabarti}}}]{adak2014rayleigh}%
  \BibitemOpen
  \bibfield  {author} {\bibinfo {author} {\bibfnamefont {A.}~\bibnamefont
  {{Adak}}}, \bibinfo {author} {\bibfnamefont {S.}~\bibnamefont {{Ghosh}}},\
  and\ \bibinfo {author} {\bibfnamefont {N.}~\bibnamefont {{Chakrabarti}}},\
  }\bibfield  {title} {\bibinfo {title} {{Rayleigh-Taylor instability in an
  equal mass plasma}},\ }\href {https://doi.org/10.1063/1.4896714} {\bibfield
  {journal} {\bibinfo  {journal} {Physics of Plasmas}\ }\textbf {\bibinfo
  {volume} {21}},\ \bibinfo {eid} {092120} (\bibinfo {year}
  {2014})}\BibitemShut {NoStop}%
\bibitem [{\citenamefont {Bera}\ \emph {et~al.}(2022)\citenamefont {Bera},
  \citenamefont {Song},\ and\ \citenamefont {Srinivasan}}]{Bera2022the}%
  \BibitemOpen
  \bibfield  {author} {\bibinfo {author} {\bibfnamefont {R.~K.}\ \bibnamefont
  {Bera}}, \bibinfo {author} {\bibfnamefont {Y.}~\bibnamefont {Song}},\ and\
  \bibinfo {author} {\bibfnamefont {B.}~\bibnamefont {Srinivasan}},\ }\bibfield
   {title} {\bibinfo {title} {The effect of viscosity and resistivity on
  rayleigh–taylor instability induced mixing in magnetized
  high-energy-density plasmas},\ }\href
  {https://doi.org/10.1017/S0022377821001343} {\bibfield  {journal} {\bibinfo
  {journal} {Journal of Plasma Physics}\ }\textbf {\bibinfo {volume} {88}},\
  \bibinfo {pages} {905880209} (\bibinfo {year} {2022})}\BibitemShut {NoStop}%
\bibitem [{\citenamefont {Khan}\ and\ \citenamefont
  {Sharma}(2023)}]{khan2023investigation}%
  \BibitemOpen
  \bibfield  {author} {\bibinfo {author} {\bibfnamefont {N.}~\bibnamefont
  {Khan}}\ and\ \bibinfo {author} {\bibfnamefont {P.}~\bibnamefont {Sharma}},\
  }\bibfield  {title} {\bibinfo {title} {Investigation of rayleigh--taylor
  instability and internal waves in strongly coupled rotating magnetized
  quantum plasma},\ }\href@noop {} {\bibfield  {journal} {\bibinfo  {journal}
  {Journal of Astrophysics and Astronomy}\ }\textbf {\bibinfo {volume} {44}},\
  \bibinfo {pages} {7} (\bibinfo {year} {2023})}\BibitemShut {NoStop}%
\bibitem [{\citenamefont {Chen}(2019)}]{chen2012introduction}%
  \BibitemOpen
  \bibfield  {author} {\bibinfo {author} {\bibfnamefont {F.~F.}\ \bibnamefont
  {Chen}},\ }\href {https://doi.org/https://doi.org/10.1007/978-3-319-22309-4}
  {\emph {\bibinfo {title} {Introduction to plasma physics and controlled
  fusion}}}\ (\bibinfo  {publisher} {Springer Cham},\ \bibinfo {year}
  {2019})\BibitemShut {NoStop}%
\bibitem [{\citenamefont {Misra}\ and\ \citenamefont
  {Abdikian}(2024)}]{misra2024transverse}%
  \BibitemOpen
  \bibfield  {author} {\bibinfo {author} {\bibfnamefont {A.~P.}\ \bibnamefont
  {Misra}}\ and\ \bibinfo {author} {\bibfnamefont {A.}~\bibnamefont
  {Abdikian}},\ }\href {https://arxiv.org/abs/2408.04404} {\bibinfo {title}
  {Transverse instability of electron-acoustic solitons in a relativistic
  degenerate astrophysical magnetoplasma}} (\bibinfo {year} {2024}),\ \Eprint
  {https://arxiv.org/abs/2408.04404} {arXiv:2408.04404 [physics.plasm-ph]}
  \BibitemShut {NoStop}%
\bibitem [{\citenamefont {Berezhiani}\ and\ \citenamefont
  {Mahajan}(1995)}]{berezhiani1995large}%
  \BibitemOpen
  \bibfield  {author} {\bibinfo {author} {\bibfnamefont {V.}~\bibnamefont
  {Berezhiani}}\ and\ \bibinfo {author} {\bibfnamefont {S.}~\bibnamefont
  {Mahajan}},\ }\bibfield  {title} {\bibinfo {title} {Large relativistic
  density pulses in electron-positron-ion plasmas},\ }\href@noop {} {\bibfield
  {journal} {\bibinfo  {journal} {Physical Review E}\ }\textbf {\bibinfo
  {volume} {52}},\ \bibinfo {pages} {1968} (\bibinfo {year}
  {1995})}\BibitemShut {NoStop}%
\bibitem [{\citenamefont {Borthakur}\ \emph {et~al.}(2018)\citenamefont
  {Borthakur}, \citenamefont {Talukdar}, \citenamefont {Neog},\ and\
  \citenamefont {Borthakur}}]{borthakur2018}%
  \BibitemOpen
  \bibfield  {author} {\bibinfo {author} {\bibfnamefont {S.}~\bibnamefont
  {Borthakur}}, \bibinfo {author} {\bibfnamefont {N.}~\bibnamefont {Talukdar}},
  \bibinfo {author} {\bibfnamefont {N.~K.}\ \bibnamefont {Neog}},\ and\
  \bibinfo {author} {\bibfnamefont {T.~K.}\ \bibnamefont {Borthakur}},\
  }\bibfield  {title} {\bibinfo {title} {Study of plasma parameters in a pulsed
  plasma accelerator using triple langmuir probe},\ }\href
  {https://doi.org/10.1063/1.5009796} {\bibfield  {journal} {\bibinfo
  {journal} {Physics of Plasmas}\ }\textbf {\bibinfo {volume} {25}},\ \bibinfo
  {pages} {013532} (\bibinfo {year} {2018})},\ \Eprint
  {https://arxiv.org/abs/https://pubs.aip.org/aip/pop/article-pdf/doi/10.1063/1.5009796/14750551/013532\_1\_online.pdf}
  {https://pubs.aip.org/aip/pop/article-pdf/doi/10.1063/1.5009796/14750551/013532\_1\_online.pdf}
  \BibitemShut {NoStop}%
\bibitem [{\citenamefont {Chandrasekhar}(1935)}]{chandrasekhar1935the}%
  \BibitemOpen
  \bibfield  {author} {\bibinfo {author} {\bibfnamefont {S.}~\bibnamefont
  {Chandrasekhar}},\ }\bibfield  {title} {\bibinfo {title} {{The Highly
  Collapsed Configurations of a Stellar Mass. (Second Paper.)}},\ }\href
  {https://doi.org/10.1093/mnras/95.3.207} {\bibfield  {journal} {\bibinfo
  {journal} {Monthly Notices of the Royal Astronomical Society}\ }\textbf
  {\bibinfo {volume} {95}},\ \bibinfo {pages} {207} (\bibinfo {year}
  {1935})}\BibitemShut {NoStop}%
\bibitem [{\citenamefont {Boshkayev}\ \emph {et~al.}(2016)\citenamefont
  {Boshkayev}, \citenamefont {Rueda}, \citenamefont {Zhami}, \citenamefont
  {Kalymova},\ and\ \citenamefont {Balgymbekov}}]{boshkayev2016equilibrium}%
  \BibitemOpen
  \bibfield  {author} {\bibinfo {author} {\bibfnamefont {K.}~\bibnamefont
  {Boshkayev}}, \bibinfo {author} {\bibfnamefont {J.~A.}\ \bibnamefont
  {Rueda}}, \bibinfo {author} {\bibfnamefont {B.}~\bibnamefont {Zhami}},
  \bibinfo {author} {\bibfnamefont {Z.~A.}\ \bibnamefont {Kalymova}},\ and\
  \bibinfo {author} {\bibfnamefont {G.~S.}\ \bibnamefont {Balgymbekov}},\
  }\bibfield  {title} {\bibinfo {title} {Equilibrium structure of white dwarfs
  at finite temperatures},\ }in\ \href@noop {} {\emph {\bibinfo {booktitle}
  {International Journal of Modern Physics: Conference Series}}},\
  Vol.~\bibinfo {volume} {41}\ (\bibinfo {organization} {World Scientific},\
  \bibinfo {year} {2016})\ p.\ \bibinfo {pages} {1660129}\BibitemShut {NoStop}%
\bibitem [{\citenamefont {Shah}\ \emph {et~al.}(2010)\citenamefont {Shah},
  \citenamefont {Qureshi},\ and\ \citenamefont {Tsintsadze}}]{shah2010effect}%
  \BibitemOpen
  \bibfield  {author} {\bibinfo {author} {\bibfnamefont {H.}~\bibnamefont
  {Shah}}, \bibinfo {author} {\bibfnamefont {M.}~\bibnamefont {Qureshi}},\ and\
  \bibinfo {author} {\bibfnamefont {N.}~\bibnamefont {Tsintsadze}},\ }\bibfield
   {title} {\bibinfo {title} {Effect of trapping in degenerate quantum
  plasmas},\ }\href@noop {} {\bibfield  {journal} {\bibinfo  {journal} {Physics
  of Plasmas}\ }\textbf {\bibinfo {volume} {17}},\ \bibinfo {pages} {032312}
  (\bibinfo {year} {2010})}\BibitemShut {NoStop}%
\bibitem [{\citenamefont {Timmes}\ and\ \citenamefont
  {Arnett}(1999)}]{timmes1999accuracy}%
  \BibitemOpen
  \bibfield  {author} {\bibinfo {author} {\bibfnamefont {F.}~\bibnamefont
  {Timmes}}\ and\ \bibinfo {author} {\bibfnamefont {D.}~\bibnamefont
  {Arnett}},\ }\bibfield  {title} {\bibinfo {title} {The accuracy, consistency,
  and speed of five equations of state for stellar hydrodynamics},\ }\href@noop
  {} {\bibfield  {journal} {\bibinfo  {journal} {The Astrophysical Journal
  Supplement Series}\ }\textbf {\bibinfo {volume} {125}},\ \bibinfo {pages}
  {277} (\bibinfo {year} {1999})}\BibitemShut {NoStop}%
\bibitem [{\citenamefont {Landau}\ and\ \citenamefont
  {Lifshitz}(2013)}]{landau2013statistical}%
  \BibitemOpen
  \bibfield  {author} {\bibinfo {author} {\bibfnamefont {L.}~\bibnamefont
  {Landau}}\ and\ \bibinfo {author} {\bibfnamefont {E.}~\bibnamefont
  {Lifshitz}},\ }\href@noop {} {\emph {\bibinfo {title} {Statistical Physics:
  Volume 5}}},\ \bibinfo {number} {v. 5}\ (\bibinfo  {publisher} {Elsevier},\
  \bibinfo {year} {2013})\BibitemShut {NoStop}%
\bibitem [{\citenamefont {Dey}\ \emph {et~al.}(2024)\citenamefont {Dey},
  \citenamefont {Banerjee}, \citenamefont {Misra},\ and\ \citenamefont
  {Bhowmik}}]{dey2024ion}%
  \BibitemOpen
  \bibfield  {author} {\bibinfo {author} {\bibfnamefont {R.}~\bibnamefont
  {Dey}}, \bibinfo {author} {\bibfnamefont {G.}~\bibnamefont {Banerjee}},
  \bibinfo {author} {\bibfnamefont {A.~P.}\ \bibnamefont {Misra}},\ and\
  \bibinfo {author} {\bibfnamefont {C.}~\bibnamefont {Bhowmik}},\ }\bibfield
  {title} {\bibinfo {title} {Ion-acoustic solitons in a relativistic fermi
  plasma at finite temperature},\ }\href@noop {} {\bibfield  {journal}
  {\bibinfo  {journal} {Scientific Reports}\ }\textbf {\bibinfo {volume}
  {14}},\ \bibinfo {pages} {26872} (\bibinfo {year} {2024})}\BibitemShut
  {NoStop}%
\end{thebibliography}%
\nopagebreak
\end{document}